\documentclass[aps,prb,reprint,groupedaddress, showpacs]{revtex4-1}
\usepackage{graphicx,graphics}
\usepackage{epstopdf}
\usepackage{dcolumn}
\usepackage{amsmath,amssymb,amsfonts}
\usepackage{latexsym,verbatim}
\usepackage{bm}
\usepackage{bbold}
\usepackage{color}
\usepackage{ulem}
\usepackage[breaklinks=true,colorlinks,citecolor=blue,linkcolor=blue,urlcolor=blue]{hyperref}
\usepackage[abs]{overpic}
\usepackage{textcomp} 
\usepackage{mathtools}
\usepackage{braket}
\usepackage{soul}    

\begin{document}

\title{Inter- and intra-band Coulomb interactions between holes in silicon nanostructures}

\author{Andrea Secchi}
\email{andrea.secchi@nano.cnr.it}
\affiliation{Centro S3, CNR-Istituto di Nanoscienze, via G. Campi 213/A, I-41125 Modena, Italy}
\author{Laura Bellentani} 
\affiliation{Centro S3, CNR-Istituto di Nanoscienze, via G. Campi 213/A, I-41125 Modena, Italy}
\author{Andrea Bertoni} 
\affiliation{Centro S3, CNR-Istituto di Nanoscienze, via G. Campi 213/A, I-41125 Modena, Italy}
\author{Filippo Troiani} 
\affiliation{Centro S3, CNR-Istituto di Nanoscienze, via G. Campi 213/A, I-41125 Modena, Italy}

\begin{abstract}
We present a full derivation of the interaction Hamiltonian for holes in silicon within the six--band envelope-function scheme, which appropriately describes the valence band close to the $\boldsymbol{\Gamma}$ point. The full structure of the single-hole eigenstates is taken into account, including the Bloch part. The scattering processes caused by the Coulomb interaction are shown to be both intraband and interband, the latter being mostly short-ranged. In the asymptotic long-range limit, the effective potential tends to the screened Coulomb potential, and becomes purely intraband, as assumed in previous models. We apply our model to compute the excitation spectra of two interacting holes in prototypical silicon quantum dots, taking into account different dielectric environments. It is shown that, in the highly screened regime, short-range interactions (both intra- and inter-band) can be very relevant, while they lose importance when there is no screening other than the one proper of the bulk silicon crystal. In the latter case, we predict the formation of hole Wigner molecules. 
\end{abstract}

\date{\today}

\maketitle

\section{Introduction}

 Silicon has played for decades an essential role in the traditional semiconductor-based information technology. More recently, it has been recognized as an excellent host material for new devices in quantum computation and spintronics. In fact, Si crystals naturally consist of 95\% non-magnetic nuclei, a percentage that can be further increased through isotopic purification\cite{Zwanenburg13}. This makes Si a candidate for the realization of quantum dot (QD) spin qubits, as the hyperfine interaction between the spin qubit and the nuclear spins of the host material typically represents the main source of decoherence and spin relaxation in other (III-V) materials \cite{Loss98,Koppens05,Petta05}. The ability to confine and control single or few electrons in Si QDs, a crucial requisite for implementing quantum computation, was achieved experimentally in the early 2010s\cite{Zwanenburg09, Simmons11, Yang12}. The values of the decoherence times achieved in Si QDs\cite{Veldhorst14, Veldhorst15, Boross16} now exceed by few orders of magnitude the demonstrated gating times \cite{Veldhorst15, Zajac18, Watson18}.

Si-based microelectronics can benefit from advanced, well-established industrial fabrication techniques\cite{Shiraki11, Sverdlov11}. This is an exceedingly important asset for achieving scalability and integration of Si qubits with control hardware. With respect to this objective, the realization and characterization of spin qubits in QDs embedded in commercially available CMOS SOI platforms offer promising perspectives\cite{Hutin16, Maurand16, DeFranceschi16, Veldhorst17, Bohuslavskyi17, Bonen19, Hutin19, Ansaloni20}. These progresses provide both a scientific and a technological motivation for the theoretical study of Si QD qubits.

The standard approach to theoretically characterize few-particle states in semiconductor nanostructures includes, as a starting point, the derivation and diagonalization of the single-particle Hamiltonian, obtained within the envelope-function approach, pioneered by  L\"uttinger and Kohn\cite{LK}. Here, the wave function is factorized into the product of a Bloch state and of an envelope function, which displays a slow spatial variation, in comparison with the lattice parameter. The envelope function is the solution of an effective Sch\"rodinger-like equation, which is determined by the external fields (confinement potentials and possibly a static magnetic field) and the effective $\boldsymbol{k} \cdot \boldsymbol{p}$ Hamiltonian \cite{Voon_book}. It is then possible to trace out the rapidly-varying Bloch states, which greatly reduces the complexity of the problem. If $M$ energy bands are relevant, with $M>1$, then the envelope functions are spinors with $M$ position-dependent components. In Si, both conduction and valley bands require, in general, a spinorial formulation. A number of crucial functionalities of spin qubits in Si depends on single-particle states, and specifically on the mixing between the bands. For example, recent works on single-hole spin qubits have thoroughly investigated the spectra and the dependence of the Larmor and Rabi frequencies on the orientation of the external magnetic field and the confinement gates, within different multi-band approaches \cite{Venitucci18, Venitucci19}.

The presence of two or more interacting particles results in a rich physics and offers further opportunities for qubit encoding, manipulation and readout. In these situations, the role of the Coulomb interaction is generally crucial. However, this is often included in theoretical models via a small number of parameters (direct and exchange interactions), which only account for intraband scattering processes \cite{Li10, Culcer10, Cota18, Kornich18}.  More comprehensive calculations are based on exact-diagonalization, or configuration-interaction (CI) procedures. These require, as input, the one-body and two-body matrix elements of the fully interacting Hamiltonian between Slater determinants built from a set of single-particle states. In the case of QD systems, the latter are generally written as the products of envelope functions and Bloch states, as mentioned above. The first main objective of this article is to derive the full interaction Hamiltonian (i.e., the two-body matrix elements of the Coulomb interaction) for Si nanostructures, such as QDs, populated by holes lying in the valence band close to the $\boldsymbol{\Gamma}$ point. Our point is that many interband scattering processes due to the Coulomb interaction exist in Si, and we provide explicit expressions and numerical values that allow to fully include them in CI calculations.

CI calculations for the case of interacting electrons in Si QDs have been presented in several works\cite{Wang10, Wang11, Raith12, Nielsen12}. In Refs.~\onlinecite{Wang10, Wang11} an accurate model is considered, related to a two-electron Si double QD, which accounts for two of the six conduction-band valleys, and the intervalley Coulomb interaction is claimed to be negligible. Even if these considerations hold for a system in a certain configuration, they cannot be generalized to arbitrary QDs or particle numbers, as the effect of different Coulomb terms depends crucially on the degree of localization of the two-particle states. In the case of holes, we find that interband terms are short-ranged and are therefore expected not to have a significant impact on states where particles are on average well separated in space (as in the ground triplet states in double QDs). In contrast, multiple occupation of a single dot implies a much smaller inter-particle distance, such that short-ranged effects can be relevant \cite{Raith12, Nielsen12}. Moreover, the spatial localization of the holes can be reduced -- and correspondingly, the impact of short-ranged interactions can be increased -- by the presence of a dielectric environment (provided, e.g., by close metallic leads) that screens the long-range Coulomb repulsion in the dot. In such cases, the interband Coulomb interaction might become one of the channels inducing band mixing, which must be taken into account very carefully in the simulation of crucial qubit operation, such as the exchange-based quantum gates or the read out based on the Pauli-blockade.

Here we focus on hole states, which are described by 4 bands (light-holes and heavy-holes), plus two additional (split-off) bands, which might be close enough in energy to be relevant, e.g., in the presence of strain. We show that Coulomb scattering induces a great variety of transitions between such bands. The situation is qualitatively different from that encountered in electronic Si QDs, where the degenerate conduction valleys are centered on different $\boldsymbol{k}$ points. As an additional motivation, we mention that analogous short-range features of the Coulomb interaction have been shown to be relevant in the case of carbon-nanotube QDs. Systematic theories\cite{Egger98, Ando06, Mayrhofer08}, experiments\cite{Pecker13}, and CI calculations\cite{Secchi13} have confirmed that the often neglected intervalley Coulomb scattering processes (which are inherently short-ranged) affect the two-electron wave functions and open additional energy gaps that cannot be explained with the intravalley Coulomb interaction only. Therefore, it is worth to investigate whether similar short-ranged processes are relevant in hole-based Si QDs. The second main objective of this article is to answer this question through CI calculations of the excitation spectra of two holes confined in Si QDs, taking into account all interaction processes. We provide a systematic study of two-hole spectra in three exemplary anisotropic Si QDs, as a function of a variable bulk dielectric constant, which mimics a variable dielectric environment. We show that, for a low screening, the computed two-hole spectra exhibit signatures of Wigner crystallization. On the other hand, when the screening of long-range interactions is high, short-range interactions become more relevant, and we quantify their impact on the two-hole spectra.

This paper is organized as follows. In Sec. \ref{sec: single} we introduce the single-hole eigenstates with the Bloch states corresponding to the $\boldsymbol{\Gamma}$ point. In Sec. \ref{sec: many} we introduce the many-hole Hamiltonian and the effective band-dependent potentials. In Sec. \ref{sec: approx} we discuss the approximations which are necessary for the derivation of the short-range and long-range effective interactions. These are obtained in Sections \ref{Sec: SR} and \ref{Sec: LR}, respectively, and collected in Sec. \ref{sec: summary W}. In Sec. \ref{sec: continuum}, we rework the formulas for the effective interactions in a way suitable for their implementation in CI codes. Finally, in Section \ref{sec: Numerical} we show and discuss the results of CI calculations of the two-hole spectra. Additional technical details related to the derivations are collected in Appendices \ref{app: Clebsch}-\ref{app: smooth}.

\section{Single-hole states at the $\Gamma$ point}
\label{sec: single}

Each unit cell in Si contains two atoms, whose positions are specified by the vectors
\begin{align}
\boldsymbol{\tau}_0 = (0,0,0) \,, \quad \boldsymbol{\tau}_1 = \frac{a}{4} (1,1,1) \,,
\end{align}
where $a = 0.5431$ nm \cite{Shiraki11} is the cubic cell edge. The lattice translation vectors are given by
\begin{align}
\boldsymbol{R} & \equiv \boldsymbol{R}(\boldsymbol{n}) \equiv   \frac{a}{2} \left( n_2 + n_3   \, , \, n_1 + n_3   \, , \, n_1 + n_2   \right)   \,,
\end{align}
for every triple of integers $\boldsymbol{n} = (n_1,n_2,n_3)$. A generic atomic position vector can then be written as $\boldsymbol{R}_j \equiv \boldsymbol{R} + \boldsymbol{\tau}_j$, with $j \in \lbrace 0, 1 \rbrace$.

We write the relevant Bloch states at the $\boldsymbol{\Gamma}$ point in tight-binding form as \cite{Chadi75, Voon_book}
\begin{align}
\big| \varepsilon^+_{\alpha, \sigma} \big> \equiv \frac{1}{\sqrt{N_{\rm c}}} 
\sum_{\boldsymbol{R}} \sum_{j} \frac{ (-1)^j }{\sqrt{2}}
\big| p_{\alpha} , \boldsymbol{R}_j \big> \otimes \big| \sigma \big> \,.
\label{Bloch at Gamma}
\end{align}
Here, $N_{\rm c}$ is the number of unit cells, $\boldsymbol{R}$ runs over their positions, $\big< \boldsymbol{r} \big|   p_{\alpha}, \boldsymbol{R}_j \big> \equiv \phi_{p_{\alpha}}(\boldsymbol{r} - \boldsymbol{R}_j)$ is an atomic orbital centered at the position $\boldsymbol{R}_j$ with the symmetry of a $p_{\alpha}$ orbital ($\alpha = x,y,z$), and $\big| \sigma \big>$ is a single-particle spinor ($\sigma = \pm 1 $). Within the shell picture, the states used in the description of the valence band at the $\boldsymbol{\Gamma}$ point are the $3 p_{\alpha}$ atomic orbitals. However, it is more convenient to adopt the Hartree-Fock orbitals\cite{Watson61}, as they allow for a better description of the chemical bonds of single-particle orbitals in a mean-field approach. 

In the presence of spin-orbit coupling, it is convenient to switch to the $(J,M)$ representation, where $J$ and $M$ are the quantum numbers associated with the square modulus and the $z$-component of a particle's total angular momentum, respectively. In particular, we include a $J = 3/2$ quartet, with $M \in \lbrace 3/2, 1/2, -1/2, -3/2 \rbrace$, and a $J = 1/2$ doublet, with $M \in \lbrace 1/2, -1/2  \rbrace$. This is accomplished via the transformation
\begin{align}
\big| \varepsilon^+_{J,M} \big> =  \sum_{\alpha, \sigma} S_{B,\alpha,\sigma} \big| \varepsilon^+_{\alpha, \sigma} \big> \,,
\label{Bloch states silicon}
\end{align}
where $B \equiv (J,M)$ and $S_{B,\alpha,\sigma}$ is the matrix of the Clebsch-Gordan coefficients \cite{Voon_book} (see Appendix \ref{app: Clebsch} for more details).

In the presence of a confinement potential that varies smoothly on the length scale of the lattice parameter, a single-hole eigenstate (labelled by an index $\nu$) can be written, in the envelope-function scheme, as
\begin{align}
  \big| \nu \big> & = \frac{1}{\sqrt{\mathcal{N}}}  \sum_B \sum_{\boldsymbol{R}} \sum_j \sum_{\alpha , \sigma}   
   (-1)^j S_{B , \alpha, \sigma}   \big| \Psi_{\nu, B, \alpha  , \boldsymbol{R}_j } \big> \! \otimes  \! \big| \sigma \big>   \,,
\label{single-particle eigenstates}
\end{align}
where 
\begin{align}
\big| \Psi_{\nu, B, \alpha , \boldsymbol{R}_j } \big> = \int d \boldsymbol{r} \, \psi_{\nu, B}(\boldsymbol{r})  \, \phi_{p_{\alpha}}(\boldsymbol{r} - \boldsymbol{R}_j) \big| \boldsymbol{r} \big> \,,
\end{align}
$\psi_{\nu, B}(\boldsymbol{r})$ is an envelope function, and the normalization constant is $\mathcal{N} = \mathcal{V}_{\rm QD} / \mathcal{V}_{\rm at}$, where $\mathcal{V}_{\rm at}$ is the volume occupied by a single atom in the Si crystal, and $\mathcal{V}_{\rm QD}$ is a normalization volume for the envelope functions, defined by
\begin{align}
\sum_B \int d \boldsymbol{r} \psi^*_{\nu', B}(\boldsymbol{r}) \, \psi_{\nu, B}(\boldsymbol{r}) = \delta_{\nu, \nu'} \, \mathcal{V}_{\rm QD} \,.
\end{align}

For a part of the following derivations, it will be useful to switch from the Cartesian to the spherical basis $\phi_m$, where $m \in \lbrace +1,0,-1 \rbrace$ is the eigenvalue of $\hat{\ell}_z$ (with $l = 1$):
\begin{align}
 \phi_{\pm 1}(\boldsymbol{r}) = 
  \frac{1}{\sqrt{2}} \Big[ \phi_{p_{x}}(\boldsymbol{r}) \pm {\rm i} \, \phi_{p_{y}}(\boldsymbol{r})\Big]  \,,\quad  \phi_{0}(\boldsymbol{r}) =   \phi_{p_{z}}(\boldsymbol{r})    \,.
\label{orbitals m}
\end{align}

\section{Many-body Hamiltonian}
\label{sec: many}

In the following, we denote with $\lbrace a \rbrace$ any set of four ordered quantities, explicitly labelled as $a_1, a_2, a_3, a_4$. For example, $\lbrace \nu \rbrace \equiv (\nu_1, \nu_2, \nu_3, \nu_4)$ and $\lbrace B \rbrace \equiv (B_1, B_2, B_3, B_4)$. 
In its diagonal form, the single-hole Hamiltonian reads
\begin{align}
\hat{H}_{\rm SH} = \sum_{\nu} E_{\nu} \hat{c}^{\dagger}_{\nu} \hat{c}_{\nu} \,, 
\end{align}
where $\nu$ labels the single-hole eigenstates. The interaction Hamiltonian has the general form
\begin{align}
\hat{H}_{\rm INT} = \frac{1}{2}\sum_{ \lbrace \nu \rbrace} V_{ \lbrace \nu \rbrace } \hat{c}^{\dagger}_{\nu_1} \hat{c}^{\dagger}_{\nu_2} \hat{c}_{\nu_3} \hat{c}_{\nu_4} \,,
\end{align}
with
\begin{align}
V_{\lbrace \nu \rbrace } & = \sum_{\sigma, \sigma'}\int {\rm d} \boldsymbol{r} \int {\rm d} \boldsymbol{r}'
\big< \nu_1 \big|   \boldsymbol{r}, \sigma \big> \,  
\big< \nu_2 \big|  \boldsymbol{r}', \sigma' \big> \nonumber \\ 
& \quad \times V\left( \boldsymbol{r} - \boldsymbol{r}' \right) \big< \boldsymbol{r}', \sigma' \big|  \nu_3 \big> \, \big< \boldsymbol{r}, \sigma \big| \nu_4 \big> \,.
\label{Coulomb integrals}
\end{align}
Here, $V\left( \boldsymbol{r} - \boldsymbol{r}' \right)$ is the screened Coulomb potential between two point charges; although we will keep our derivation general with respect to the choice of the interaction potential, in Appendix \ref{app: screened potential} we discuss the details of the Vinsome-Richardson expression \cite{Vinsome71, Richardson71}, which is suitable for Si. At the vertices of the two-particle interaction processes (positions $\boldsymbol{r}$ and $\boldsymbol{r}'$), the spin components $\sigma$ and $\sigma'$ are conserved. However, at each vertex the interaction can induce transitions between different bands, i.e. different values of $B$. To see this, we rewrite Eq.~\eqref{Coulomb integrals} using the explicit forms of the single-hole eigenstates given in Eq.~\eqref{single-particle eigenstates}:
\begin{align}
V_{ \lbrace \nu \rbrace}  
& =  \sum_{ \lbrace B \rbrace }     \int {\rm d} \boldsymbol{r} \int {\rm d} \boldsymbol{r}'
 \psi^*_{\nu_1, B_1}(\boldsymbol{r})   \,   
 \psi^*_{\nu_2, B_2}(\boldsymbol{r}')  \nonumber \\
 & \quad \times   W_{ \lbrace B \rbrace }\left( \boldsymbol{r} - \boldsymbol{r}' \right) \, 
 \psi_{\nu_3, B_3}(\boldsymbol{r}')   \,  
 \psi_{\nu_4, B_4}(\boldsymbol{r})  \,.
 \label{band dependent V}
\end{align}
Here, we have introduced the effective band-dependent interaction potential,
\begin{align}
W_{\lbrace B \rbrace }\left( \boldsymbol{r} - \boldsymbol{r}' \right) & \equiv  V\left( \boldsymbol{r} - \boldsymbol{r}' \right) \frac{1}{ \mathcal{N}^2 }  \sum_{ \lbrace \boldsymbol{R} \rbrace } \sum_{ \lbrace j \rbrace } (-1)^{j_1+j_2+j_3+j_4} \nonumber \\
& \quad \times \sum_{ \lbrace m \rbrace}    
   F_{B_1 , B_4}^{m_1, m_4} F_{B_2 , B_3}^{m_2, m_3} \nonumber \\
& \quad \times   \phi^*_{m_1}(\boldsymbol{r} - \boldsymbol{R}_{1, j_1}) \, \phi^*_{m_2}(\boldsymbol{r}' - \boldsymbol{R}_{2, j_2})  \nonumber \\
& \quad \times \phi_{m_3}(\boldsymbol{r}' - \boldsymbol{R}_{3, j_3}) \,  \phi_{m_4}(\boldsymbol{r} - \boldsymbol{R}_{4, j_4}) \,.
\label{W before localization}
\end{align}
The matrix $F_{B, B'}^{m,m'}$ is given explicitly in Appendix \ref{app: Clebsch}, together with the details of the transformation. Since $F_{B, B'}^{m,m'} \neq 0$ for $B \neq B'$, interband scattering processes are possible.

In its current form, Eq.~\eqref{W before localization} is of no practical use, as it involves an excessively demanding quadruple summation over all the $N_{\rm a}= 2 N_{\rm c}$ atoms in the crystal ($N^4_{\rm a}$ terms), not to mention the summations over the other indices. The aim of this work is to transform this expression into one that can be more easily implemented and used in practical calculations.

\section{Approximations on the effective interaction potential}
\label{sec: approx}

We now resume the derivation of the multi-band interaction potential, and proceed with the manipulation of Eq.~\eqref{W before localization}.

\subsection{Two-center integral approximation}

The main difficulty associated with the calculation of the Coulomb interaction potential arises from the presence of orbitals centered at 4 different atomic sites. As a result, the Coulomb matrix elements [Eq.~\eqref{band dependent V}] are given by 4-centre integrals. A widely used approximation\cite{Ando06, Mayrhofer08} consists in keeping only the one- and two-center integrals, where  
\begin{align}
   \boldsymbol{R}_{1,j_1}   =  \boldsymbol{R}_{4, j_4} \quad {\rm and} \quad  \boldsymbol{R}_{2,j_2}   =  \boldsymbol{R}_{3, j_3} \,,
 \label{TB approximation}
\end{align}
and discarding the three- and four-center ones. The rationale for this approximation is that the orbitals decay exponentially with the distance from their center: therefore, the leading terms in Eq.~\eqref{W before localization} are expected to be those where the two orbitals involving the same hole coordinate are centered on the same site. We shall also adopt this approximation, which can be justified {\it a posteriori} by the fact that the asymptotic limit of the interaction potential coincides with the screened Coulomb potential (Section \ref{sec: summary W}). A possible route to go beyond this approximation is sketched in Appendix \ref{app: Rudenberg}, but remains essentially beyond the scope of the present work. 

\subsection{Slow spatial dependence of the envelope functions}

We now consider the full matrix element of the hole-hole interaction [Eq.~\eqref{band dependent V}]. After applying the two-center integral approximation, this reads as 
\begin{align}
V_{\lbrace \nu \rbrace } & \approx  \sum_{\lbrace B \rbrace}      \sum_{ \lbrace m \rbrace }    
  F^{\, m_1, \, m_4}_{\, B_1 ,  \, B_4 }  \,  F^{\, m_2, \, m_3}_{\, B_2 ,  \, B_3 }  \nonumber \\
  & \quad \times \frac{1}{ \mathcal{N}^2 }  \sum_{\boldsymbol{R}_j  , \boldsymbol{R}'_{j'}} \int {\rm d} \boldsymbol{r} \int {\rm d} \boldsymbol{r}'
 \psi^*_{\nu_1, B_1}(\boldsymbol{r})   \,   
 \psi^*_{\nu_2, B_2}(\boldsymbol{r}')  \nonumber \\
 & \quad \times 
 \psi_{\nu_3, B_3}(\boldsymbol{r}')   \,  
 \psi_{\nu_4, B_4}(\boldsymbol{r})   V\left( \boldsymbol{r} - \boldsymbol{r}' \right) \,  \phi^*_{m_1}(\boldsymbol{r} - \boldsymbol{R}_j )    \nonumber \\
& \quad \times   \phi^*_{m_2}(\boldsymbol{r}' - \boldsymbol{R}'_{j'})   \, \phi_{m_3}(\boldsymbol{r}' - \boldsymbol{R}'_{j'}) \,  \phi_{m_4}(\boldsymbol{r} - \boldsymbol{R}_j) \,.
\label{slow}
\end{align}

We then exploit the slow variation of the envelope functions on the length scale of the lattice parameter, combined with the strong localization of the atomic orbitals. If the envelope function is practically constant over the volume occupied by an atom, one has that 
\begin{align}
\psi_{\nu, B}(\boldsymbol{r}) \phi_{m}(\boldsymbol{r} - \boldsymbol{R}_j) \simeq \psi_{\nu, B}(\boldsymbol{R}_j) \phi_{m}(\boldsymbol{r} - \boldsymbol{R}_j) \,.
\label{envelope approx}
\end{align}
Under this approximation, the four envelope functions drop out of the integrals over $\boldsymbol{r}$ and $\boldsymbol{r}'$, and thus
\begin{align}
V_{\lbrace \nu \rbrace } & \approx \frac{1}{ \mathcal{N}^2 }  \sum_{\boldsymbol{R}_j  , \boldsymbol{R}'_{j'}} 
 \sum_{ \lbrace B \rbrace }    \psi^*_{\nu_1, B_1}(\boldsymbol{R}_j) \, \psi^*_{\nu_2, B_2}(\boldsymbol{R}'_{j'})   \nonumber \\
 & \quad \times   
  \psi_{\nu_3, B_3}(\boldsymbol{R}'_{j'})   \,  
 \psi_{\nu_4, B_4}(\boldsymbol{R}_j)       W_{\lbrace B \rbrace}(\boldsymbol{R}_j, \boldsymbol{R}'_{j'} )  \,,
 \label{V before continuum}
\end{align}
where 
\begin{align}
W_{\lbrace B \rbrace}(\boldsymbol{R}_j, \boldsymbol{R}'_{j'} ) & \equiv \sum_{ \lbrace m \rbrace}    
F^{\, m_1, \, m_4}_{\, B_1 ,  \, B_4 }  \,  F^{\, m_2, \, m_3}_{\, B_2 ,  \, B_3 }  \nonumber \\ 
& \quad \times \int {\rm d} \boldsymbol{r}_1 \int {\rm d} \boldsymbol{r}_2  \, \phi^*_{m_1}(\boldsymbol{r}_1 ) \, \phi^*_{m_2}(\boldsymbol{r}_2) \nonumber \\
& \quad \times  V\left( \boldsymbol{r}_1 - \boldsymbol{r}_2 + \boldsymbol{R}_j - \boldsymbol{R}'_{j'}\right)   \nonumber \\
& \quad \times \phi_{m_3}(\boldsymbol{r}_2) \,  \phi_{m_4}(\boldsymbol{r}_1) \,.
\label{W}
\end{align}
Although the integrals extend over the whole space, the domain over which the integrand is non-zero is a small neighbourhood of the origin ($\boldsymbol{r}_1 = \boldsymbol{r}_2 = \boldsymbol{0}$), because of the localization of the atomic orbitals. Therefore, in the relevant domain, $|\boldsymbol{r}_1 - \boldsymbol{r}_2|$ is of the order of the linear size of the unit cell, and one can distinguish two regimes:
\begin{itemize}
\item The short-range regime, where $\boldsymbol{R}_j = \boldsymbol{R}'_{j'}$.
\item The long-range regime, where $\boldsymbol{R}_j \neq \boldsymbol{R}'_{j'}$, and one can assume that $\left| \boldsymbol{R}_j - \boldsymbol{R}'_{j'} \right| \gg \left| \boldsymbol{r}_1 - \boldsymbol{r}_2 \right|$. 
\end{itemize}
These two regimes are treated in Sections \ref{Sec: SR} and \ref{Sec: LR}, respectively.

\section{Short-range effective interaction}
\label{Sec: SR}

In the short-range (SR) case, the expression of the effective interaction [Eq.~\eqref{W}, with $\boldsymbol{R}_j = \boldsymbol{R}'_{j'}$] becomes
\begin{align}
  W^{\rm SR}_{\lbrace B \rbrace }   
  \equiv 
  \sum_{ \lbrace m \rbrace }    
F^{\, m_1, \, m_4}_{\, B_1 ,  \, B_4 }  \,  F^{\, m_2, \, m_3}_{\, B_2 ,  \, B_3 }  U_{\lbrace m \rbrace}   \,,
\label{W SR}
\end{align}
where 
\begin{align}
U_{\lbrace m \rbrace} & \equiv   \int {\rm d} \boldsymbol{r}_1 \int {\rm d} \boldsymbol{r}_2  \, \phi^*_{ m_1 }(\boldsymbol{r}_1 ) \, \phi^*_{m_2}(\boldsymbol{r}_2) \, V\left( \boldsymbol{r}_1 - \boldsymbol{r}_2 \right)  \nonumber \\
& \quad \times \phi_{ m_3 }(\boldsymbol{r}_2) \,  \phi_{ m_4 }(\boldsymbol{r}_1) \,.
\label{Hubbard parameters}
\end{align}
Hereafter, we compute the Hubbard parameters $U_{\lbrace m \rbrace}$ in the approximation
\begin{align}
V\left( \boldsymbol{r}_1 - \boldsymbol{r}_2 \right) \simeq V_{\rm C}\left( \boldsymbol{r}_1 - \boldsymbol{r}_2 \right) \,
\label{SR approx on V}.
\end{align}
This is justified by the fact that the integrand vanishes when $\left| \boldsymbol{r}_1 - \boldsymbol{r}_2 \right|$ is large with respect to the size of the orbitals, while the screening is negligible in the opposite limit, which gives the major contribution to the integral.

We note that there are 81 Hubbard parameters $U_{\lbrace m \rbrace}$. However, most of them are identically zero, and the remaining ones are related by several symmetry relations, which greatly reduce the number of independent quantities to be evaluated.

\subsection{Evaluation of the Hubbard parameters}

The first step in the calculation of the Hubbard parameters is to write the orbitals in spherical coordinates: 
\begin{align}
\phi_{m}(\boldsymbol{r}_i) = R_{3,1}(r_i) \, Y_{1,m}(\theta_i, \varphi_i) \,, \quad i \in \lbrace 1,2\rbrace \,,
\end{align}
where $R_{n,l}$ and $Y_{l,m}$ are the radial orbital function and the spherical harmonic, respectively, taken for $n=3$ and $l=1$, which is the case of interest.

Next, we expand the unscreened Coulomb potential [see Eq.~\eqref{SR approx on V} and the related discussion] in the series of Legendre polynomials $P_{\ell}(\cos\omega) \equiv P_{\ell, 0}(\cos\omega)$. In CGS units,
\begin{align}
    V_{\rm C}(| \boldsymbol{r}_1 - \boldsymbol{r}_2 |) = \frac{e^2}{| \boldsymbol{r}_1 - \boldsymbol{r}_2 |} = e^2\sum_{\ell=0}^{+\infty}\frac{r_{<}^\ell}{r_>^{\ell+1}}P_{\ell}(\cos\omega) \,,
    \label{Legendre expansion}
\end{align}
where $r_< = \min(r_1,r_2)$, $r_> = \max(r_1,r_2)$, and $\omega$ is the angle between $\boldsymbol{r}_1$ and $\boldsymbol{r}_2$. The angle $\omega$ can be written as a function of $\theta_1$, $\theta_2$, $\varphi_1$, and $\varphi_2$, using the spherical harmonic addition theorem\cite{Griffith}. This allows us to perform the integrals over the solid angles in Eq.~\eqref{Hubbard parameters} and to obtain, after some algebra:
\begin{align}
    U_{\lbrace m \rbrace }   & =   \delta_{m_1 , m_4} \delta_{m_2,m_3}     \nonumber \\
    & \quad \times \left[ F_0 +        \frac{(-1)^{|m_1| + |m_2|} \left( 2 - |m_1| \right) \left( 2 - |m_2| \right) }{25}   F_2  \right]   \nonumber \\
    & \quad   +  \delta_{m_1+m_2,m_3+m_4} (1 - \delta_{m_1,m_4}) (1 - \delta_{m_2,m_3})  \nonumber \\
    & \quad \times    \frac{3 \sqrt{  \left(|m_1| + |m_4|\right) \left(|m_2| + |m_3|\right) } }{25}    \,   F_2     \,,
    \label{Hubbard in terms of F}
\end{align}
where
\begin{align}
& F_0   = e^2 \int_0^{\infty} dr_1 r_1^2 \int_0^{\infty} dr_2 r_2^2 \,  \frac{1}{r_>} \, R^2_{3,1}(r_1)R^2_{3,1}(r_2) \,, \nonumber \\
& F_2   = e^2 \int_0^{\infty} dr_1 r_1^2 \int_0^{\infty} dr_2 r_2^2 \,  \frac{r^2_<}{r^3_>}R^2_{3,1}(r_1)R^2_{3,1}(r_2) 
\label{Slater-Condon}
\end{align} 
are Slater-Condon parameters \cite{Slater29, Condon30}, depending on the radial wave function associated with the $\phi_m$ orbitals. The full derivation leading from Eq.~\eqref{Hubbard parameters} to Eq.~\eqref{Hubbard in terms of F} is presented in Appendix \ref{app: Hubbard}.

One can show that, out of the 81 Hubbard parameters corresponding to the different values of $(m_1,m_2,m_3,m_4)$, only the following 19 are different from zero:
\begin{align}
& U_{0, 0, 0, 0} = F_0    + \frac{4}{25} F_2  \,,     \nonumber \\
& U_{\pm 1, \pm 1, \pm 1, \pm 1} =  U_{\pm 1, \mp 1, \mp 1, \pm 1}
= F_0    + \frac{1}{25} F_2 \,, \nonumber \\
& U_{\pm 1, 0, 0, \pm 1} = U_{0, \pm 1, \pm 1, 0} = F_0    - \frac{2}{25} F_2 \,, 
\nonumber \\   
& U_{\pm 1, \mp 1, \pm 1, \mp 1} = \frac{6}{25}  F_2 \,,   \nonumber \\    
& U_{0, 0, \pm 1, \mp 1}  =   U_{\pm 1, \mp 1, 0, 0} =  U_{\pm 1, 0, \pm 1, 0} = U_{0,\pm 1, 0,\pm 1} \nonumber \\
& \quad \quad \quad \quad \,\, = \frac{3}{25}  F_2 \,.
\label{independent Hubbard}
\end{align}

The problem is now reduced to the determination of the Slater-Condon parameters $F_0 \equiv F_0(3p,3p)$ and $F_2 \equiv F_2(3p,3p)$. These quantities depend on the radial orbital wave functions [see Eq.~\eqref{Slater-Condon}], which are sensitive to the electronic configuration. The quantities $F_0$ and $F_2$ can be computed analytically, e.g. using Hartree-Fock radial wave functions\cite{Fisk68, Watson61}. The calculation presented in Ref.~\onlinecite{Watson61} yields 
\begin{align}
F_0 = 8.99037 \,\, {\rm eV} \,, \quad  F_2 = 4.53941 \,\, {\rm eV} \,.
\label{numerical values}
\end{align} 

\subsection{Short-range potential in terms of the Hubbard parameters}
\label{sec: Hub}
 
Hereafter, we proceed to perform the sums appearing in Eq.~\eqref{W SR}, using Eqs.~\eqref{independent Hubbard}, and state the results. The full derivation is presented in Appendix \ref{app: short-range}. 

The terms that contribute to the short-range interaction potentials can be divided in three classes:
\begin{align}
  W^{\rm SR}_{\lbrace B \rbrace } &  = W^{\rm SR, \, intra}_{\lbrace B \rbrace }  + W^{\rm SR, \, part}_{\lbrace B \rbrace } + W^{\rm SR, \, inter}_{\lbrace B \rbrace }  \,.
\label{WSR separated}
\end{align} 
The first class is formed by 36 fully intraband terms, characterized by $B_1 = B_4$ and $B_2 = B_3$:
\begin{align}
W^{\rm SR, \, intra}_{\lbrace B \rbrace } 
  =   \delta_{B_1,B_4}  \delta_{B_2,B_3} U^{\rm intra}_{B_1, B_2} \,,
 \label{WSR fully intra}
\end{align}
where
\begin{align}
U^{\rm intra}_{B_1, B_2} =     F_0  + \delta_{J_1, \frac{3}{2}} \delta_{J_2, \frac{3}{2}}  (-1)^{| M_1| -  | M_2|}   F_2^{\star}     \, ,
 \label{USR fully intra}
\end{align}
where $F_2^{\star} \equiv F_2/25$. The second class is formed by 32 partially intraband terms, characterized by $B_1 = B_4$ and $B_2 \neq B_3$, or $B_2 = B_3$ and $B_1 \neq B_4$:
\begin{align} 
W^{\rm SR, \, part}_{\lbrace B \rbrace }   =   \delta_{B_1,B_4} U^{\rm part}_{B_1; B_2, B_3}         +  \delta_{B_2,B_3} U^{\rm part}_{B_2; B_1, B_4}  \,,
\label{WSR partial}
\end{align}
where
\begin{align} 
U^{\rm part}_{B_1; B_2, B_3} & =       \delta_{J_1, \frac{3}{2}} \left( 1 - \delta_{J_2, J_3} \right) \delta_{M_2,M_3} \delta_{| M_2| ,  \frac{1}{2}}  \nonumber \\
& \quad \times  (-1)^{|M_1| + \frac{1}{2}}      \sqrt{2}  F^{\star}_2    \,.
\label{USR partial}
\end{align}
The third class includes 120 fully interband terms, characterized by $B_1 \neq B_4$ and $B_2 \neq B_3$:
\begin{align}
   W^{\rm SR, \, inter}_{\lbrace B \rbrace } 
   =     U^{(1),\, \rm inter}_{\lbrace B \rbrace } + U^{(2),\,\rm inter}_{B_1, B_4; B_2, B_3}   + U^{(2),\,\rm inter}_{B_2, B_3; B_1, B_4}   \,,
 \label{WSR inter}
\end{align}
where
\begin{align}
 U^{(1),\, \rm inter}_{\lbrace B  \rbrace} & \equiv \delta_{  M_1 ,   M_4 }  \delta_{ | M_1 |  ,   \frac{1}{2} } \, \delta_{  M_2 ,   M_3 } \delta_{ | M_2 |  ,   \frac{1}{2} }  \nonumber \\
 & \quad \times \left( 1 - \delta_{J_1, J_4} \right) \left( 1 - \delta_{J_2, J_3} \right) 2 F^{\star}_2  \,,
\label{USR inter 1}
\end{align}
and
\begin{align}
& U^{(2),\,\rm inter}_{B_1, B_4; B_2, B_3} \nonumber \\
& \equiv    \Big[ Y_{J_1}   \,  \delta_{J_4, \frac{3}{2} } \,   \delta_{M_1, - \frac{1}{2}}  \, \delta_{M_4, - \frac{3}{2}}       +  \delta_{J_1, \frac{3}{2} }  \,  Y_{J_4}  \, \delta_{M_1,  \frac{3 }{2}} \,     \delta_{M_4,   \frac{1}{2}}  \nonumber \\
& \quad       + \left( J_1 - J_4 \right)  \delta_{M_1,   \frac{ 1}{2}}   \,      \delta_{M_4,  - \frac{ 1}{2}} \Big] \nonumber \\
& \quad \times   \Big[ \delta_{J_2, \frac{3}{2} } \, Y_{J_3}  \, \delta_{M_2, - \frac{3}{2}}  \,     \delta_{M_3, - \frac{1}{2}}        +  Y_{J_2}  \, \delta_{J_3, \frac{3}{2} } \,     \delta_{M_2,   \frac{1}{2}}  \,   \delta_{M_3,  \frac{3 }{2}}   \nonumber \\
& \quad       + \left( J_3 - J_2 \right) \delta_{M_2,  - \frac{ 1}{2}} \, \delta_{M_3,   \frac{ 1}{2}}     \Big]  3  F^{\star}_2 \nonumber \\ 
& \quad + \Big( X_{J_1} \,    \delta_{J_4, \frac{3}{2} } \, \delta_{M_1,   \frac{1}{2}} \, \delta_{M_4,  - \frac{3 }{2}} -  \delta_{J_1, \frac{3}{2} } \,  X_{J_4} \, \delta_{M_1, \frac{3}{2}} \, \delta_{M_4, - \frac{1}{2}}          \Big) \nonumber \\
& \quad \times \Big( \delta_{J_2, \frac{3}{2} } \, X_{J_3}  \, \delta_{M_2, - \frac{3}{2}}  \, \delta_{M_3,   \frac{1}{2}}          -   X_{J_2} \,   \delta_{J_3, \frac{3}{2} } \, \delta_{M_2, - \frac{1}{2}}  \, \delta_{M_3,    \frac{3 }{2}}  \Big) \nonumber \\
& \quad \times 6 F_2^{\star} \,.
\label{USR inter 2}
\end{align}
Equations \eqref{USR inter 1} and \eqref{USR inter 2} give all the non-vanishing interband parameters entering Eq.~\eqref{WSR inter}. These are listed in Tables \ref{tab:WSR 1}, \ref{tab:WSR 2}, and \ref{tab:WSR 3}, and classified according to the values of  $\lbrace J \rbrace$. 

\begin{table}
 \begin{tabular}{| c   | c c c c || c |  } 
 \hline 
 $2 \lbrace J \rbrace  $ & $2M_1$ & $2M_2$ & $2M_3$ & $2M_4$ & $U^{\rm inter}_{\lbrace B \rbrace}$   \\ [0.5ex] 
 \hline 
 $(3, 3, 3, 3)$ & 
 $3s$ & $t$ & $3s$ & $t$ &   $2F^{\star}_2$   \\
&  $t$ & $3 s$ & $t$ & $3 s$ & $2F^{\star}_2$   \\ 
&  $3s$ & $-3s$ & $-s t$ & $s t$ & $t 2F^{\star}_2$    \\
&  $st$ & $-st$ & $-3s$ & $3s$ & $t 2 F^{\star}_2$  \\ 
 \hline
\end{tabular}\caption{Interband Hubbard parameters $U^{\rm inter}_{\lbrace B \rbrace}$, in the cases where there is no transfer of $J$ at both interaction vertices ($J_1 - J_4 = J_2 - J_3 = 0$), valid for any $s = \pm 1$ and $t= \pm 1$.} \label{tab:WSR 1}
\end{table}

\begin{table}
 \begin{tabular}{| c | c c c c || c |   } 
 \hline 
 $2 \lbrace J \rbrace$   & $2M_1$ & $2M_2$ & $2M_3$ & $2M_4$ & $U^{\rm inter}_{\lbrace B \rbrace}$   \\ [0.5ex] 
 \hline 
 $(3, 3, 3, 1)$ & 
 $3s$ & $-s$ & $3s$ & $-s$ &   $2\sqrt{2}F^{\star}_2$   \\ 
&   $3s$ & $-3s$ & $s$ & $-s$ & $-2\sqrt{2}F^{\star}_2$   \\
&  $s$ & $s$ & $3s$ & $-s$ & $-s \sqrt{6} F^{\star}_2$  \\
&  $-s$ & $3s$ & $s$ & $s$ & $s \sqrt{6}F^{\star}_2$   \\
&  $3s$ & $-3s$ & $-s$ & $s$ & $-\sqrt{2}F^{\star}_2$    \\
&  $3s$ & $s$ & $3s$ & $s$ & $-\sqrt{2}F^{\star}_2$   \\
 \hline
 $(3, 3, 1, 3)$ & 
 $3s$ & $-3s$ & $s$ & $-s$ &   $-2\sqrt{2}F^{\star}_2$    \\
&   $s$ & $-3s$ & $s$ & $-3s$ & $2\sqrt{2}F^{\star}_2$   \\
&  $3s$ & $-s$ & $s$ & $s$ & $s \sqrt{6}F^{\star}_2$    \\
&  $s$ & $s$ & $-s$ & $3s$ & $-s\sqrt{6}F^{\star}_2$   \\
&  $3s$ & $-3s$ & $-s$ & $s$ & $-\sqrt{2}F^{\star}_2$    \\
&  $s$ & $3s$ & $s$ & $3s$ & $-\sqrt{2}F^{\star}_2$   \\
 \hline
 $(3, 1, 3, 3)$ & 
 $3s$ & $-s$ & $3s$ & $-s$ &   $2\sqrt{2}F^{\star}_2$    \\
&   $s$ & $-s$ & $3s$ & $-3s$ & $-2\sqrt{2}F^{\star}_2$   \\
&  $3s$ & $-s$ & $s$ & $s$ & $-s\sqrt{6}F^{\star}_2$   \\
&  $s$ & $s$ & $-s$ & $3s$ & $s \sqrt{6}F^{\star}_2$   \\
&  $3s$ & $s$ & $3s$ & $s$ & $-\sqrt{2}F^{\star}_2$   \\
&  $-s$ & $s$ & $3s$ & $-3s$ & $-\sqrt{2}F^{\star}_2$   \\
 \hline
 $(1, 3, 3, 3)$ & 
 $s$ & $-3s$ & $s$ & $-3s$ &   $2\sqrt{2}F^{\star}_2$    \\
&   $s$ & $-s$ & $3s$ & $-3s$ & $-2\sqrt{2}F^{\star}_2$   \\ 
&  $s$ & $s$ & $3s$ & $-s$ & $s \sqrt{6}F^{\star}_2$   \\
&  $-s$ & $3s$ & $s$ & $s$ & $-s\sqrt{6}F^{\star}_2$   \\
&  $s$ & $3s$ & $s$ & $3s$ & $-\sqrt{2}F^{\star}_2$   \\
&  $-s$ & $s$ & $3s$ & $-3s$ & $-\sqrt{2}F^{\star}_2$   \\ 
 \hline
\end{tabular}\caption{Interband Hubbard parameters $U^{\rm inter}_{\lbrace B \rbrace}$, in the cases where there is transfer of $J$ at only one of the two interaction vertices ($|J_1 - J_4| = 1$ and $J_2 = J_3$, or $J_1 = J_4$ and $|J_2 - J_3| = 1$), valid for any $s = \pm 1$.} \label{tab:WSR 2}
\end{table}

\begin{table}
 \begin{tabular}{| c | c c c c || c | } 
 \hline 
 $2 \lbrace J \rbrace$   & $2M_1$ & $2M_2$ & $2M_3$ & $2M_4$ & $U^{\rm inter}_{\lbrace B \rbrace}$  \\ [0.5ex] 
 \hline 
 $(3, 3, 1, 1)$ & 
 $3s$ & $-3s$ & $s$ & $-s$ &   $-4F^{\star}_2$   \\
&  $s$ & $-s$ & $s$ & $-s$ & $-3F^{\star}_2$   \\
&  $-s$ & $3s$ & $s$ & $s$ & $-s\sqrt{3}F^{\star}_2$   \\
&  $3s$ & $-s$ & $s$ & $s$ & $-s\sqrt{3}F^{\star}_2$   \\
&  $3s$ & $-3s$ & $-s$ & $s$ & $F^{\star}_2$ \\
&  $s$ & $t$ & $t$ & $s$ & $2F^{\star}_2$   \\
 \hline
 $(3, 1, 3, 1)$ & 
 $3s$ & $-s$ & $3s$ & $-s$ &   $4F^{\star}_2$   \\
&  $s$ & $-s$ & $s$ & $-s$ & $3F^{\star}_2$   \\  
&  $s$ & $s$ & $3s$ & $-s$ & $s\sqrt{3}F^{\star}_2$   \\
&  $3s$ & $-s$ & $s$ & $s$ & $s\sqrt{3}F^{\star}_2$   \\ 
&  $3s$ & $s$ & $3s$ & $s$ & $F^{\star}_2$   \\
&  $s$ & $t$ & $t$ & $s$ & $2F^{\star}_2$   \\
 \hline
 $(1, 3, 1, 3)$ & 
 $-s$ & $3s$ & $-s$ & $3s$ &   $4F^{\star}_2$    \\
&  $s$ & $-s$ & $s$ & $-s$ & $3F^{\star}_2$    \\
&   $-s$ & $3s$ & $s$ & $s$ & $s \sqrt{3}F^{\star}_2$   \\
&  $s$ & $s$ & $-s$ & $3s$ & $s \sqrt{3}F^{\star}_2$   \\
&  $s$ & $3s$ & $s$ & $3s$ & $F^{\star}_2$   \\
&  $s$ & $t$ & $t$ & $s$ & $2F^{\star}_2$   \\
 \hline
 $(1, 1, 3, 3)$ & 
 $-s$ & $s$ & $-3s$ & $3s$ & $-4F^{\star}_2$    \\
&  $s$ & $-s$ & $s$ & $-s$ & $-3F^{\star}_2$   \\ 
&  $s$ & $s$ & $3s$ & $-s$ & $-s\sqrt{3}F^{\star}_2$   \\
&  $s$ & $s$ & $-s$ & $3s$ & $-s\sqrt{3}F^{\star}_2$   \\
&  $s$ & $-s$ & $-3s$ & $3s$ & $F^{\star}_2$   \\ 
&  $s$ & $t$ & $t$ & $s$ & $2F^{\star}_2$   \\
 \hline
\end{tabular}\caption{Interband Hubbard parameters $U^{\rm inter}_{\lbrace B \rbrace}$, in the cases where there is transfer of $J$ at both interaction vertices ($|J_1 - J_4| = |J_2 - J_3| = 1$), valid for any $s=\pm1$ and $t=\pm1$.} \label{tab:WSR 3}
\end{table}

\section{Long-range effective interaction}
\label{Sec: LR}

We now consider the effective interaction [Eq.~\eqref{W}] in the long-range (LR) regime, where $\boldsymbol{R}_j \neq \boldsymbol{R}'_{j'}$ and $\left| \boldsymbol{R}_j - \boldsymbol{R}'_{j'} \right| \gg \left| \boldsymbol{r}_1 - \boldsymbol{r}_2 \right|$. In this case, the expansion of the interaction potential in Taylor series gives:
\begin{align}
& V\left( \boldsymbol{r}_1 - \boldsymbol{r}_2 + \boldsymbol{R}_j - \boldsymbol{R}'_{j'}\right) \nonumber \\
& \approx V\left( \boldsymbol{R}_j - \boldsymbol{R}'_{j'}\right) + 
\sum_{\alpha} \left( \alpha_1 - \alpha_2 \right) \partial_{\alpha} V\left(   \boldsymbol{R}_j - \boldsymbol{R}'_{j'}\right) \nonumber \\
& \quad + \frac{1}{2} \sum_{\alpha, \beta} \left( \alpha_{1} - \alpha_2 \right) \left( \beta_{1} - \beta_2 \right) \partial^2_{\alpha, \beta}V\left(   \boldsymbol{R}_j - \boldsymbol{R}'_{j'}\right) \,,
\label{Taylor}
\end{align}
where $\alpha, \beta \in \lbrace x,y,z\rbrace$, and $\partial_{\alpha}V(\boldsymbol{R}) \equiv \frac{\partial V(\boldsymbol{R})}{\partial R_{\alpha}} $. When the expansion Eq.~\eqref{Taylor} is substituted into Eq.~\eqref{W}, three terms are obtained, for $\boldsymbol{R}_j \neq \boldsymbol{R}'_{j'}$:
\begin{align}
  W^{\rm LR}_{\lbrace B \rbrace}(\boldsymbol{R}_j, \boldsymbol{R}'_{j'} )  
& \simeq \sum_{n=0}^2 W^{{\rm LR,} (n)}_{\lbrace B \rbrace }(\boldsymbol{R}_j, \boldsymbol{R}'_{j'} )  \,.
\label{W LR three terms}
\end{align}
In the remainder of this Section, we use the shorthand $\boldsymbol{R} \equiv \boldsymbol{R}_j - \boldsymbol{R}'_{j'}  \equiv   R (C_x, C_y, C_z)$, where $R = \left| \boldsymbol{R} \right|$, and $C^2_x + C^2_y+ C^2_z = 1$. 

\subsection{Long-range potential, zeroth-order}

The zeroth-order term from Eq.~\eqref{W LR three terms} is 
\begin{align}
 W^{\rm LR, (0)}_{\lbrace B \rbrace}(\boldsymbol{R} )  & = \sum_{\lbrace m \rbrace}    
F^{\, m_1, \, m_4}_{\, B_1 ,  \, B_4 }  \,  F^{\, m_2, \, m_3}_{\, B_2 ,  \, B_3 } \int {\rm d} \boldsymbol{r} \, \phi^*_{m_1}(\boldsymbol{r} )  \,  \phi_{m_4}(\boldsymbol{r})    \nonumber \\
 & \quad \times  \int {\rm d} \boldsymbol{r}'   \, \phi^*_{m_2}(\boldsymbol{r}') \,  \phi_{m_3}(\boldsymbol{r}')  \, V\left( \boldsymbol{R} \right)  \nonumber \\
& =    V\left( \boldsymbol{R} \right) \delta_{B_1, B_4} \delta_{B_2, B_3}  \,,
\label{W LR}
\end{align}
where we have used the orthogonality of the orbitals,
\begin{align}
\int {\rm d} \boldsymbol{r} \, \phi^*_{m}(\boldsymbol{r}  )  \,  \phi_{m'}(\boldsymbol{r} ) = \delta_{m,m'} \,,
\label{ortho}
\end{align}
as well as the trace property of the matrix $F$ [see Eq.~\eqref{trace F matrix} in Appendix \ref{app: Clebsch}]. We note that this term of the LR interaction is fully intraband.

\subsection{Long-range potential, first-order}

The first-order term from Eq.~\eqref{W LR three terms} is 
\begin{align}
 W^{\rm LR, (1)}_{\lbrace B \rbrace }(\boldsymbol{R} )  
& =   \sum_{\alpha}\partial_{\alpha} V\left(   \boldsymbol{R} \right)   \sum_{\lbrace m \rbrace}    
F^{\, m_1, \, m_4}_{\, B_1 ,  \, B_4 }   F^{\, m_2, \, m_3}_{\, B_2 ,  \, B_3 }   \nonumber \\
& \quad \times  \int {\rm d} \boldsymbol{r}_1 \int {\rm d} \boldsymbol{r}_2  \, \phi^*_{m_1}(\boldsymbol{r}_1 ) \, \phi^*_{m_2}(\boldsymbol{r}_2)      \nonumber \\
& \quad \times \left( \alpha_1 - \alpha_2 \right) \phi_{m_3}(\boldsymbol{r}_2) \,  \phi_{m_4}(\boldsymbol{r}_1) \nonumber \\
& =    \sum_{\alpha} \partial_{\alpha}V\left(   \boldsymbol{R} \right)    \sum_{m , m'}  \int {\rm d} \boldsymbol{r}   \, \phi^*_{m }(\boldsymbol{r}  ) \,     \alpha    \,   \phi_{m'}(\boldsymbol{r} )  \nonumber \\
& \quad \times  \left( \delta_{  B_2 ,    B_3 }     
F^{\, m , \, m'}_{\, B_1 ,  \, B_4 }     -   \delta_{ B_1 ,   B_4 }      
   F^{\, m , \, m'}_{\, B_2 ,  \, B_3 }           \right)    \,,
\label{W LR 1}
\end{align}
where we have used Eq.~\eqref{ortho}. The integrals appearing in Eq.~\eqref{W LR 1} vanish,
\begin{align}
 \int {\rm d} \boldsymbol{r}   \, \phi^*_{m }(\boldsymbol{r}  ) \,     \alpha      \,   \phi_{m'}(\boldsymbol{r} )  
 = 0 \,, \quad \forall \alpha \in \lbrace x,y,z \rbrace \,,
 \label{vanishing}
\end{align}
therefore
\begin{align}
W^{\rm LR, (1)}_{\lbrace B \rbrace}(\boldsymbol{R}  )  = 0 \,.
\end{align} 

Equation \eqref{vanishing} can be proved by observing that a product $\phi^*_{p_{\alpha'}}(\boldsymbol{r}  ) \,     \alpha \, \phi_{p_{\alpha''}}(\boldsymbol{r} )$ is always odd in one or three Cartesian coordinates, therefore $\int {\rm d} \boldsymbol{r}   \, \phi^*_{p_{\alpha'}}(\boldsymbol{r}  ) \,     \alpha \, \phi_{p_{\alpha''}}(\boldsymbol{r} ) = 0 $. Then, since the orbitals $\phi_{m }(\boldsymbol{r}  )$ appearing in Eq.~\eqref{vanishing} are linear combinations of the orbitals $\phi_{p_{\alpha}}(\boldsymbol{r} ) $ [see Eq.~\eqref{orbitals m}], the quantity in the right-hand side of Eq.~\eqref{vanishing} can be written as a linear combination of integrals of the form $\int {\rm d} \boldsymbol{r}   \, \phi^*_{p_{\alpha'}}(\boldsymbol{r}  ) \,     \alpha \, \phi_{p_{\alpha''}}(\boldsymbol{r} )$, therefore it vanishes.

\subsection{Long-range potential, second-order}
\label{sec: LR2}

The second-order term from Eq.~\eqref{W LR three terms} is
\begin{align}
& W^{\rm LR, (2)}_{\lbrace B \rbrace}(\boldsymbol{R} ) \nonumber \\
& =   \frac{1}{2} \sum_{\alpha, \beta}   \partial^2_{\alpha, \beta}V\left(   \boldsymbol{R} \right) \sum_{\lbrace m \rbrace}    
F^{\, m_1, \, m_4}_{\, B_1 ,  \, B_4 }  F^{\, m_2, \, m_3}_{\, B_2 ,  \, B_3 } \nonumber \\
& \quad \times     \int {\rm d} \boldsymbol{r}_1 \int {\rm d} \boldsymbol{r}_2  \, \phi^*_{m_1}(\boldsymbol{r}_1 ) \, \phi^*_{m_2}(\boldsymbol{r}_2) \,  \left( \alpha_{1} - \alpha_2 \right)  \nonumber \\
& \quad \times \left( \beta_{1} - \beta_2 \right)       \phi_{m_3}(\boldsymbol{r}_2)  \, \phi_{m_4}(\boldsymbol{r}_1)    \nonumber \\
& =   \frac{1}{2} \sum_{\alpha, \beta}   \partial^2_{\alpha, \beta}V\left(   \boldsymbol{R} \right)  \sum_{m , m' }   \int {\rm d} \boldsymbol{r}   \, \phi^*_{m }(\boldsymbol{r}  ) \,    \alpha  \beta     \,  \phi_{m'}(\boldsymbol{r} )   \nonumber \\
& \quad \times 
\left( \delta_{B_2 ,   B_3 } F^{  m ,   m'}_{  B_1 ,    B_4 }               +   \delta_{  B_1 ,  B_4 }        F^{  m ,   m'}_{  B_2 ,   B_3 }     \right)   \,,
\label{W LR 2}
\end{align}
where we have used Eqs.~\eqref{ortho}-\eqref{vanishing}. The expressions of the $\alpha\beta$-integrals, 
\begin{align}
& \int {\rm d} \boldsymbol{r}   \, \phi^*_{m }(\boldsymbol{r}  ) \,     \alpha \beta     \,   \phi_{m'}(\boldsymbol{r} )    \,,
\label{alphabeta integrals before}
\end{align}
for $\alpha \beta \in \lbrace x^2, y^2, z^2, x y , y z , z x \rbrace$, are provided in Appendix \ref{app: alphabeta}. As becomes apparent after switching to spherical coordinates, they are all proportional to the following quantity:
\begin{align}
\left< r^2_{3,1} \right> \equiv \int_0^{\infty} {\rm d} r r^4 \left| R_{3,1}(r) \right|^2          \,.
\label{eq:intrad2}
\end{align}

In order to compute Eq.~\eqref{eq:intrad2}, one needs to specify the radial wave function $R_{3,1}(r)$. Two alternative possibilities are considered in Appendix \ref{app: r2}: one is based on hydrogen-like orbitals with a screened nuclear charge $Z^{\star}$, and the other one on the Hartree-Fock orbitals that were used in Ref.~\onlinecite{Watson61} to obtain the values of $F_0$ and $F_2$ given in Eqs.~\eqref{numerical values}. In the first case, we first determine $Z^{\star}$ that fits Eqs.~\eqref{numerical values}, and use the resulting hydrogen-like orbital to compute Eq.~\eqref{eq:intrad2}. The two numerical results for $\left< r^2_{3,1} \right>$ are very close, differing by less than $6\%$ despite the difference in the functional forms of the radial wave functions; their average value is $\left< r^2_{3,1} \right> \approx 0.0245$ nm$^2$. 

After inserting the expressions of the $\alpha\beta$-integrals into Eq.~\eqref{W LR 2} and performing some algebraic manipulation, one gets:
\begin{align}
W^{\rm LR, (2)}_{\lbrace B \rbrace}(\boldsymbol{R}  ) & \equiv   V(\boldsymbol{R})   \Big[ \delta_{  B_1 ,  B_4 }        \delta_{B_2, B_3}   \Delta^{(2)}_{B_1,B_2}(\boldsymbol{R}) \nonumber \\
& \quad + \delta_{B_1, B_4}   \Lambda^{(2)}_{B_2, B_3}(\boldsymbol{R})    + \delta_{B_2, B_3}   \Lambda^{(2)}_{B_1, B_4}(\boldsymbol{R})  \Big] \,,
\label{W LR (2) structure}
\end{align}
where
\begin{align}
   \Delta^{(2)}_{B, B'}(\boldsymbol{R}) & \equiv  \frac{ \left< r^2_{3,1} \right>  }{ 5  V(\boldsymbol{R}) }   \Big(                 
  \Gamma^{I}_{B, B'}  \nabla^2       +       
   \Gamma^{II}_{B, B'} \, \partial^2_{z, z}    \Big) V(\boldsymbol{R}) \,,
\label{W LR 2 intraband general}
\\
\Lambda^{(2)}_{B, B'}(\boldsymbol{R}  ) & =  \frac{ \left< r^2_{3,1} \right>  }{ 5 V(\boldsymbol{R})  }   \Bigg\{                 
 \Upsilon_{B, B'} \Bigg( \frac{\nabla^2  }{3} -    \partial^2_{z, z}  \Bigg)           \nonumber \\
& \quad + \Xi^{+}_{B, B'} \frac{1}{2}   \left( \partial_{x}  -  {\rm i}   \partial_{y} \right)^2        +   \Xi^{-}_{B, B'}  \frac{1}{2}   \left( \partial_{x} + {\rm i}   \partial_{y} \right)^2        \nonumber \\
& \quad +  \Theta^{+}_{B, B'} \,  \partial_z \left(   \partial_{x}  -  {\rm i}    \partial_{y} \right)      \nonumber \\
& \quad  + \Theta^{-}_{B, B'}  \, \partial_z     \left(  \partial_{x}  +   {\rm i}    \partial_{y}   \right)    \Bigg\} V(\boldsymbol{R}) \,.
\label{W LR 2 partially intraband general}
\end{align}
The functions $\Gamma^{I}_{B, B'}$, $\Gamma^{II}_{B, B'}$, $\Upsilon_{B, B'}$, $\Xi^{\pm}_{B, B'}$, and $\Theta^{\pm}_{B, B'}$ provide selection rules and weights for the various processes. Specifically, the functions $\Gamma_{B, B'}^{I}$ and $\Gamma_{B, B'}^{II}$ [Table \ref{tab:WLR intra}] enter the definition of $\Delta^{(2)}_{B, B'}(\boldsymbol{R})$ and are therefore related to intraband scattering processes. The functions $\Upsilon_{B, B'}$, $\Xi_{B, B'}^{\pm}$ and $\Theta_{B, B'}^{\pm}$ [Table \ref{tab:WLR part}], instead, enter the definition of $\Lambda^{(2)}_{B, B'}(\boldsymbol{R})$ and are therefore related to partially intraband scattering processes.  

For a screened interaction potential of the form $V(r)=V_{\rm C}(r) /  \epsilon(r)$, one has  
\begin{align}
 \partial^2_{\alpha, \beta} V(R)  =  V(R) \left[ \mathcal{L}^{-2}(R) \,  C_{\alpha} C_{\beta}   - \delta_{\alpha, \beta} \mathcal{M}^{-2}(R)  \right]   \,, 
\end{align}
where the quantities $\mathcal{L}^{-2}(R)$ and $\mathcal{M}^{-2}(R)$ both have the dimensions of an inverse length squared, and are given by 
\begin{align}
\mathcal{L}^{-2}(R) \equiv   \frac{3}{R^2}  
+ \frac{ 3 \epsilon'(R)}{R \epsilon(R)} 
+ 2 \left[ \frac{ \epsilon'(R)}{  \epsilon(R)} \right]^2
- \frac{\epsilon''(R)}{\epsilon(R)}    \,,
\label{Lm2 general}
\end{align}
and
\begin{align}
\mathcal{M}^{-2}(R) \equiv  \frac{1}{R^2} + 
\frac{\epsilon'(R)}{R \epsilon(R)} \,.
\label{Mm2 general}
\end{align}
In Appendix \ref{app: screened potential} we show the form taken by Eqs.~\eqref{Lm2 general} and \eqref{Mm2 general} in the case of the Vinsome-Richardson potential, already mentioned in Section \ref{sec: many}. We finally obtain the explicit formulas
\begin{align}
  \Delta^{(2)}_{B,B'}(\boldsymbol{R}) & \equiv  \frac{ \left< r^2_{3,1} \right>  }{ 5   }     \Bigg\{      \left[ \mathcal{L}^{-2}(R) - 3 \mathcal{M}^{-2}(R) \right]        
   \Gamma^{I}_{B , B' }     \nonumber \\
& \quad +    \left[ \mathcal{L}^{-2}(R) \,  C_{z}^2   - \mathcal{M}^{-2}(R)  \right]
  \Gamma^{II}_{B, B'}        \Bigg\}   \,,
  \label{W LR 2 intra}     
\end{align} 
\begin{align}
&   \Lambda^{(2)}_{B, B'}(\boldsymbol{R})  \nonumber \\
& =   \frac{ \left< r^2_{3,1} \right>  }{ 5   }    \, \mathcal{L}^{-2}(R)   \Bigg[                
\left(  \frac{1}{3}    -  C^2_z  \right)   \Upsilon_{B, B'}         \nonumber \\  
& \quad +  \frac{1}{2}   \left( C_x - {\rm i} C_y \right)^2         \Xi^{+}_{B, B'}    
        + \frac{1}{2}      \left( C_x + {\rm i} C_y \right)^2       \Xi^{-}_{B, B'}        \nonumber \\
& \quad +   C_{z} \left( C_x - {\rm i} C_y \right)      \Theta^{+}_{B, B'}    
     +     C_{z} \left( C_x + {\rm i} C_y \right)       \Theta^{-}_{B, B'}     \Bigg]    \,.
\label{W LR 2 inter}  
\end{align}

\begin{table}
 \begin{tabular}{| c | c  || c | c | } 
 \hline 
 $ 2\left(J, \left| M \right| \right)  $ & $ 2\left(J', \left| M' \right| \right)$ &    $\Gamma^{I}_{B, B'}$  & $\Gamma^{II}_{B, B'}$  \\ [0.5ex] 
 \hline 
 $(3, 3)$ & $(3, 3)$ &  $2$  & $-1$     \\
 $(3, 3)$ & $(3, 1)$ &  $5/3$  &  $0$ \\
 $(3, 3)$ & $(1, 1)$ & $ 11/6$ & $-1/2    $ \\
 \hline
 $(3, 1)$ & $(3, 3)$ &  $5/3$  & $0$ \\
 $(3, 1)$ & $(3, 1)$ &   $4/3$  & $1$ \\
 $(3, 1)$ & $(1, 1)$  &  $3/2$ &  $1/2 $    \\
 \hline 
 $(1, 1)$ & $(3, 3)$ & $11/6$  &  $-1/2$ \\
 $(1, 1)$ & $(3, 1)$ &  $3/2$ &  $1/2$  \\ 
 $(1, 1)$ & $(1, 1)$ &  $5/3$ &  $0$  \\ 
 \hline  
\end{tabular}\caption{Characteristic functions for the second-order corrections to the long-range intraband scattering processes. 
} \label{tab:WLR intra}
\end{table}

\begin{table}
 \begin{tabular}{| c | c  || c ||  c | c ||  c | c | } 
 \hline 
 $ 2(J, M)  $ & $ 2(J', M')$ & $\Upsilon_{B,B'}$ & $\Xi^{+}_{B, B'}$ & $\Xi^{-}_{B, B'}$ & $\Theta^{+}_{B, B'}$  & $\Theta^{-}_{B, B'}$  \\ [0.5ex] 
 \hline 
 $(3, 3)$ & $(3, 1)$ & $0$ & $0$ & $0$ &  $ - \sqrt{ \frac{1}{3}}$ &   $0$    \\
 $(3, 3)$ & $(3, -1)$ & $0$ & $- \sqrt{\frac{1}{3}}$ & $0$ &  $0$ &   $0$    \\
 $(3, 3)$ & $(1, 1)$ & $0$& $0$ & $0$ &   $\sqrt{\frac{1}{6}}$ &   $0$    \\
 $(3, 3)$ & $(1, -1)$ & $0$ & $- \sqrt{ \frac{2}{3} }$ & $0$ &   $0$ &   $0$    \\
 \hline
 $(3, 1)$ & $(3, 3)$ & $0$ & $0$ & $0$ &  $0$   &  $- \sqrt{\frac{1}{3}}$     \\
 $(3, 1)$ & $(3, -3)$ & $0$ & $\sqrt{\frac{1}{3}}$ & $0$ & $0$   &  $0$     \\
 $(3, 1)$ & $(1, 1)$  & $\sqrt{\frac{1}{2}}$ & $0$ & $0$ &  $0$   &  $0$     \\
 $(3, 1)$ & $(1, -1)$ & $0$ & $0$ & $0$ &  $ \sqrt{\frac{1}{2}}$   &  $0$      \\
 \hline
 $(3, -1)$ & $(3, 3)$ & $0$ & $0$ & $- \sqrt{\frac{1}{3}} $ &   $0$  &  $0$  \\
 $(3, -1)$ & $(3, -3)$ & $0$ & $0$ & $0$ &  $- \sqrt{\frac{1}{3}}$  &  $0$  \\
 $(3, -1)$ & $(1, 1)$ & $0$ & $0$ & $0$ &  $0$  &  $- \sqrt{\frac{1}{2}}$     \\
 $(3, -1)$ & $(1, -1)$ & $\sqrt{\frac{1}{2}}$ & $0$ & $0$ &  $0$  &  $0$     \\
 \hline 
 $(3, -3)$ & $(3, 1)$ & $0$ & $0$ & $\sqrt{\frac{1}{3}}$ &  $0$ & $0$   \\
 $(3, -3)$ & $(3, -1)$ & $0$ & $0$ & $0$ & $0$ & $ - \sqrt{\frac{1}{3}} $    \\
 $(3, -3)$ & $(1, 1)$  & $0$ & $0$ & $ \sqrt{ \frac{2}{3} } $ &  $0$ & $0$   \\
 $(3, -3)$ & $(1, -1)$ & $0$ & $0$ & $0$ & $0$ & $ \sqrt{\frac{1}{6}}$        \\
 \hline 
 $(1, 1)$ & $(3, 3)$ & $0$ & $0$ & $0$ & $0$ &  $\sqrt{\frac{1}{6}}$   \\
 $(1, 1)$ & $(3, 1)$ & $\sqrt{\frac{1}{2}}$ &  $0$ & $0$ & $0$ & $0$     \\
 $(1, 1)$ & $(3, -1)$ & $0$ & $0$ & $0$ & $- \sqrt{\frac{1}{2}}$ & $0$     \\
 $(1, 1)$ & $(3, -3)$ & $0$ & $\sqrt{ \frac{2}{3} }$ & $0$ &  $0$ & $0$    \\
 \hline 
 $(1, -1)$ & $(3, 3)$ & $0$ & $0$ & $- \sqrt{ \frac{2}{3} } $ & $0$  &  $0$   \\
 $(1, -1)$ & $(3, 1)$ & $0$ & $0$ & $0$ & $0$  &  $ \sqrt{\frac{1}{2}} $      \\
 $(1, -1)$ & $(3, -1)$ & $\sqrt{\frac{1}{2}}$ & $0$ & $0$ & $0$  &  $0$      \\
 $(1, -1)$ & $(3, -3)$ & $0$ & $0$ & $0$ &  $\sqrt{\frac{1}{6}}$ & $0$     \\
 \hline  
\end{tabular}\caption{Characteristic functions for the long-range partially intraband scattering processes, displayed for the values of $B$ and $B'$ such that at least one among the five functions does not vanish.} \label{tab:WLR part}
\end{table}

\section{Total interaction potentials}
\label{sec: summary W}

We now summarize our findings and show the total expressions for the band-dependent interaction potentials, classified on the basis of the (non-) conservation of the band indices at the interaction vertices.

The fully intraband potential has both SR and LR components, 
\begin{align}
& W_{B, B', B', B}(\boldsymbol{R}_j, \boldsymbol{R}'_{j'} ) \nonumber \\
& \approx   \delta_{\boldsymbol{R}_j , \boldsymbol{R}'_{j'}} U^{\rm intra}_{B, B'} + \left( 1 - \delta_{\boldsymbol{R}_j , \boldsymbol{R}'_{j'}} \right) \, V\left( \boldsymbol{R}_j - \boldsymbol{R}'_{j'}\right)  \nonumber \\
& \quad \times \left[ 1 + \Delta^{(2)}_{B, B'}\left( \boldsymbol{R}_j - \boldsymbol{R}'_{j'}\right)  \right]   \,.
\label{intraband W}
\end{align}
The parameters $U^{\rm intra}_{B, B'}$, defining 36 short-ranged intraband processes in Eq.~\eqref{intraband W}, are given in Eq.~\eqref{USR fully intra}. The function $\Delta^{(2)}_{B, B'}\left( \boldsymbol{R}_j - \boldsymbol{R}'_{j'}\right)$ is given by Eq.~\eqref{W LR 2 intra}.

The partially intraband potential also exhibits both SR and LR components, 
\begin{align}
& W_{B, B', B'', B}(\boldsymbol{R}_j, \boldsymbol{R}'_{j'} ) = W_{B', B, B, B''}(\boldsymbol{R}_j, \boldsymbol{R}'_{j'} ) \nonumber \\
& \approx   \delta_{\boldsymbol{R}_j , \boldsymbol{R}'_{j'}} U^{\rm part}_{B; B', B''}   \nonumber \\
& \quad + \left( 1 - \delta_{\boldsymbol{R}_j , \boldsymbol{R}'_{j'}} \right) \, V\!\left( \boldsymbol{R}_j - \boldsymbol{R}'_{j'}\right)     \Lambda^{(2)}_{B', B''}\!\left( \boldsymbol{R}_j - \boldsymbol{R}'_{j'}\right)     \,.
\label{partially intraband W}
\end{align}
The parameters $U^{\rm part}_{B; B', B''}$, determining the 32 partially intraband processes in Eq.~\eqref{partially intraband W}, are given in Eq.~\eqref{USR partial}. The function $\Lambda^{(2)}_{B, B'}\left( \boldsymbol{R}_j - \boldsymbol{R}'_{j'}\right)$ is given by Eq.~\eqref{W LR 2 inter}.

The interband potential is completely SR, and is given by Eq.~\eqref{WSR inter}, which we rewrite here for completeness ($\boldsymbol{R}_j = \boldsymbol{R}'_{j'}$),
\begin{align}
   W^{\rm  inter}_{\lbrace B \rbrace }  
 & =  U^{(1),\, \rm inter}_{\lbrace B \rbrace } + U^{(2),\,\rm inter}_{B_1, B_4; B_2, B_3}  
 + U^{(2),\,\rm inter}_{B_2, B_3; B_1, B_4} \,.
 \label{interband W}
\end{align}
The 120 non-vanishing parameters $U^{\rm inter}_{\lbrace B \rbrace }$ satisfy the conditions $B_1 \neq B_4 $ and $ B_2 \neq B_3$, and they are synthetically listed in the formulas \eqref{USR inter 1} and \eqref{USR inter 2}.

From Eqs.~\eqref{Lm2 general} and \eqref{Mm2 general} we notice that, for $R \rightarrow \infty$,
\begin{align}
\mathcal{L}^{-2}(R) \approx \frac{3}{R^2} \,, \quad  \mathcal{M}^{-2}(R) \approx \frac{1}{R^2} \,,
\end{align} 
since the dielectric function asymptotically approaches the constant value $\epsilon_0 \equiv \lim_{R \rightarrow \infty} \epsilon(R)$. It follows that the interaction potential becomes asymptotically intraband and equal to the screened Coulomb potential:
\begin{align}
\lim_{R \rightarrow \infty} W_{\lbrace B \rbrace}(\boldsymbol{R} ) \approx \delta_{B_1, B_4} \delta_{B_2, B_3} V(R)   \,,
\label{a posteriori}
\end{align}
as the second-order corrections decay quicker with the distance $R$, namely as $\approx V(R)/ R^2$.

\section{The continuum limit }
\label{sec: continuum}

\subsection{Method}

We now restore the continuum representation for the envelope functions and the interaction potentials, by taking the continuum limit of Eq.~\eqref{V before continuum}, which can be rewritten exactly as:
\begin{align}
V_{\lbrace \nu \rbrace } & =  \sum_{\lbrace B \rbrace }          \int \frac{ d \boldsymbol{r}}{\mathcal{V}_{\rm QD}}    \int \frac{ d \boldsymbol{r}'}{\mathcal{V}_{\rm QD}}   \psi^*_{\nu_1, B_1}(\boldsymbol{r})   \, \psi^*_{\nu_2, B_2}(\boldsymbol{r}')  \, 
 \psi_{\nu_3, B_3}(\boldsymbol{r}')  \nonumber \\
& \quad \times 
  \psi_{\nu_4, B_4}(\boldsymbol{r})   \,  \widetilde{W}_{\lbrace B \rbrace}(\boldsymbol{r} , \boldsymbol{r}'  )  \,, 
\end{align}
having introduced the effective potential
\begin{align}
\widetilde{W}_{\lbrace B \rbrace}(\boldsymbol{r} , \boldsymbol{r}'  ) & \equiv \frac{1}{ \rho^2 } \sum_{\boldsymbol{R}_j  , \boldsymbol{R}'_{j'}}  \delta(\boldsymbol{r} - \boldsymbol{R}_j) \, \delta(\boldsymbol{r}' - \boldsymbol{R}'_{j'}) \nonumber \\
& \quad \times W_{\lbrace B \rbrace}(\boldsymbol{R}_j, \boldsymbol{R}'_{j'} ) \,,
\label{W continuum to be done}
\end{align}
and the nuclear density $ \rho \equiv 1 / \mathcal{V}_{\rm at}$.

We now notice that, according to our findings summarized in Section \ref{sec: summary W}, the total interaction potential $W$ can be partitioned as
\begin{align}
 W_{\lbrace B \rbrace}(\boldsymbol{R}_j, \boldsymbol{R}'_{j'})    
& \equiv \delta_{\boldsymbol{R}_j, \boldsymbol{R}'_{j'}} W^{\rm SR}_{ \lbrace B \rbrace } \nonumber \\
& \quad + \left( 1 - \delta_{\boldsymbol{R}_j, \boldsymbol{R}'_{j'}} \right) W^{\rm LR}_{\lbrace B \rbrace}(  \boldsymbol{R}_j - \boldsymbol{R}'_{j'}) \,.
\label{W partition}
\end{align} 
Combining Eq.~\eqref{W partition} with Eq.~\eqref{W continuum to be done}, one obtains
\begin{align}
\widetilde{W}_{\lbrace B \rbrace}(\boldsymbol{r} , \boldsymbol{r}'  ) & =  W^{\rm SR}_{ \lbrace B \rbrace } \delta(\boldsymbol{r} - \boldsymbol{r}') \, \frac{1}{ \rho^2 } \sum_{\boldsymbol{R}_j  }  \delta(\boldsymbol{r} - \boldsymbol{R}_j)   \nonumber \\
& \quad + W^{\rm LR}_{\lbrace B \rbrace}( \boldsymbol{r}  -  \boldsymbol{r}'   )  \, \frac{1}{ \rho^2 }  \sum_{\boldsymbol{R}_j   } \delta(\boldsymbol{r} - \boldsymbol{R}_j)  \nonumber \\
& \quad \times \sum_{  \boldsymbol{R}'_{j'} \neq \boldsymbol{0} }    \delta(  \boldsymbol{r} - \boldsymbol{r}'     - \boldsymbol{R}'_{j'})     \,.
\label{W continuum - step 1}
\end{align}

In order to perform the summations over the atomic coordinates, we replace the $\delta$-functions by smooth functions $g$, satisfying the condition 
\begin{align}
\int d \boldsymbol{r} \, g(\boldsymbol{r} - \boldsymbol{R}_j) = 1 \,.
\label{equal 1}
\end{align}
This replacement is valid because of the slow variation of the envelope functions with respect to the scale of the lattice parameter \cite{Ando06}. The definition of the functions $g$ is subjected to a certain degree of arbitrariness; a rigorous way to introduce them is the following.

We define a set of cubes $\mathcal{C}_{\boldsymbol{R}_j}$, centered on $\boldsymbol{R}_j$ and of edge $\lambda$, such that every atom $\boldsymbol{R}_j$ is the only occupier of the cube $\mathcal{C}_{\boldsymbol{R}_j}$. The cubes either are disjointed, or they share sets of points having zero volume, and their union does not necessarily cover the whole space. They are merely introduced as a way to spread the weight of a $\delta$ function over a domain of finite size. In fact, the function $g(\boldsymbol{r} - \boldsymbol{R}_j)$ is then required to have the properties
\begin{align}
\int_{\mathcal{C}_{\boldsymbol{R}_j}} d \boldsymbol{r} \, g(\boldsymbol{r} - \boldsymbol{R}_j) = 1 \,, \quad g(\boldsymbol{r} - \boldsymbol{R}_j) = 0 \quad {\rm if} \,\, \boldsymbol{r} \notin \mathcal{C}_{\boldsymbol{R}_j} \,.
\end{align}
Any function satisfying these constraints represents a suitable definition of $g$. We show a concrete solution in Appendix \ref{app: smooth}. 
 
The $g$ functions are then used to evaluate the following sums, relevant for Eq.~\eqref{W continuum - step 1}:
\begin{align}
F(\boldsymbol{r}) = \sum_{\boldsymbol{R}_j} g(\boldsymbol{r} - \boldsymbol{R}_j) \,,
 \quad
G(\boldsymbol{r}) = \sum_{\boldsymbol{R}_j \neq \boldsymbol{0}} g(\boldsymbol{r} - \boldsymbol{R}_j) \,.
\end{align}
They are related by
\begin{align}
G(\boldsymbol{r}) = F(\boldsymbol{r}) - g(\boldsymbol{r} ) \,,
\end{align}
and it holds that
\begin{align}
  G(\boldsymbol{r})     \equiv   \left\{ \begin{matrix}  0 & \, {\rm if } \, \boldsymbol{r} \in \mathcal{C}_{\boldsymbol{0}} \\
F(\boldsymbol{r}) & \, {\rm if } \, \boldsymbol{r} \notin \mathcal{C}_{\boldsymbol{0}} \end{matrix} \right.        \,.
\label{G total}
\end{align}
Besides, we notice that the average value of $F(\boldsymbol{r})$ over the crystal volume $\mathcal{V}$ is
\begin{align}
\frac{1}{\mathcal{V}} \int  d \boldsymbol{r} \, F(\boldsymbol{r}  ) = \frac{ N_{\rm a} }{\mathcal{V}} = \rho \,,
\label{average F}
\end{align}
independently of the size of the cube $\lambda^3$.

Although the replacement of the $\delta$ with the $g$ functions yields computable quantities, computationally demanding summations over all the lattice positions are still required. In order to make the problem tractable, we replace the true Si lattice with an equally spaced grid, having the same density. This is expected to have no significant consequences on the evaluation of $V_{\lbrace \nu \rbrace}$ in the continuum limit, due to the slow spatial dependence of the envelope functions. The grid is defined by the vectors $\boldsymbol{R}_{\boldsymbol{n}} = \lambda (n_x, n_y, n_z)$, where $\lambda$ is chosen such that the volume $\lambda^3$ of the cube $\mathcal{C}_{\boldsymbol{R}_{\boldsymbol{n}}}$ is the same as half the volume of the unit cell of the Si lattice, i.e. $\lambda = a / 2$ and $\rho = 1 / \lambda^3$. In this situation, the cubes introduced above cover the whole space, and each of them shares a face with a neighbour. Let us now focus on the cube centered in $\boldsymbol{R}_{\boldsymbol{0}} = \boldsymbol{0}$, and on its nearest, next-nearest, and next-next-nearest neighbours. The union of these 27 cubes forms a larger cube, which we denote as $\mathcal{R}$, with edge equal to $3 \lambda$.

In the continuum limit, the function $F(\boldsymbol{r})$ is replaced with its average value $\rho$ in all the grid cells not belonging to $\mathcal{R}$. There, we leave $F(\boldsymbol{r}) = g(\boldsymbol{r})$ and $G(\boldsymbol{r}) = 0$ in the cube at the origin, and we modify the values of $F(\boldsymbol{r})$ in the other 26 singled-out cubes in such a way that it evolves continuously to the average value $\rho$ [see Eq.~\eqref{average F}] at the borders of $\mathcal{R}$, while keeping the correct integral properties of the $\delta$ functions. After the replacement
\begin{align}
F(\boldsymbol{r}) \rightarrow \widetilde{F}(\boldsymbol{r}) \quad {\rm if} \,\, \boldsymbol{r} \in \mathcal{R} \setminus \mathcal{C}_{\boldsymbol{0}} \,,
\end{align}
we proceed to determine $\widetilde{F}(\boldsymbol{r}) $. As in the case of the determination of $g$, there is a degree of arbitrariness in the definition of $\widetilde{F}(\boldsymbol{r})$; an explicit solution is shown in Appendix \ref{app: smooth}.

Going back to Eq.~\eqref{W continuum - step 1} and using the smooth functions and the related concepts introduced in the previous Section, one has that
\begin{align}
& \sum_{\boldsymbol{R}_j  }  \delta(\boldsymbol{r} - \boldsymbol{R}_j) \equiv F(\boldsymbol{r}) \approx \rho \,, \nonumber \\
& \delta(\boldsymbol{r} - \boldsymbol{r}') \approx g(\boldsymbol{r} - \boldsymbol{r}') \,, \nonumber \\
& \sum_{  \boldsymbol{R}'_{j'} \neq \boldsymbol{0} }    \delta(  \boldsymbol{r} - \boldsymbol{r}'     - \boldsymbol{R}'_{j'}) \equiv G(\boldsymbol{r} - \boldsymbol{r}') \,.
\end{align}
Therefore $\widetilde{W}_{\lbrace B \rbrace}(\boldsymbol{r} , \boldsymbol{r}'  ) \rightarrow \widetilde{W}_{\lbrace B \rbrace}(\boldsymbol{r} - \boldsymbol{r}'  )$ depends only on the difference of the hole coordinates, and
\begin{align}
\widetilde{W}_{\lbrace B \rbrace}(\boldsymbol{r}    )   \equiv   W^{\rm SR}_{ \lbrace B \rbrace } g_d(\boldsymbol{r} )      + W^{\rm LR}_{\lbrace B \rbrace}( \boldsymbol{r}    )       \, G_d(  \boldsymbol{r}   )       \,.
\label{W continuum - step 2}
\end{align}
where we have introduced the dimensionless functions $g_d(\boldsymbol{r}) \equiv g(\boldsymbol{r}) / \rho$ and $G_d(\boldsymbol{r}) \equiv G(\boldsymbol{r}) / \rho$, whose explicit expressions are given in Appendix \ref{app: smooth}.

\subsection{The band-dependent potentials}

We now discuss and plot the various types of band-dependent potentials in the continuum limit, starting from the results collected in Section \ref{sec: summary W}. All the plots presented here are done using the values of $F_0$ and $F_2$ given in Eq.~\eqref{numerical values}.

The fully intraband potentials read as
\begin{align}
 \widetilde{W}_{B, B', B', B}(\boldsymbol{r}  ) & =     g_d(\boldsymbol{r}  )  \, U^{\rm intra}_{B, B'}   \nonumber \\
& \quad   +        G_d(  \boldsymbol{r}  )  \,  V\left( \boldsymbol{r}    \right)    \left[ 1 + \Delta^{(2)}_{B, B'}\left( \boldsymbol{r}    \right)  \right]    \,.
\label{intraband W continuum}
\end{align}
and are plotted in Fig.~\ref{Fig: W intra, SR} (short-range) and Fig.~\ref{Fig: W intra, LR} (long-range). It can be seen that the difference between distinct short-range intraband potentials is most pronounced close to $\boldsymbol{r}=\boldsymbol{0}$. In the long-range regime, the potentials are weakly dependent on the values of $B$ and $B'$, due to the second-order long-range corrections, displayed separately in Fig.~\ref{Fig: W intra, second order}. The splitting occurs on a short distance scale ($\approx 0.5$ nm for the chosen direction), due to the quick decay of $\Delta^{(2)}_{B,B'}$. All the long-range intraband potentials converge to the screened Coulomb potential (right-hand side of Fig.~\ref{Fig: W intra, LR}).

The partially intraband potentials read as
\begin{align}
& \widetilde{W}_{B, B', B'', B}(\boldsymbol{r}   )   = \widetilde{W}_{B', B, B, B''}(\boldsymbol{r}    )     \nonumber \\
& = g_d(\boldsymbol{r}  ) \,  U^{\rm part}_{ B; B', B'' }    +    G_d(  \boldsymbol{r}    ) \,    V(\boldsymbol{r}    ) \, \Lambda^{(2)}_{B', B''}(\boldsymbol{r}   )            \,.
\label{partially intraband W continuum}
\end{align}
Their short- and long-range parts are plotted in Fig.~\ref{Fig: W part, SR} and Fig.~\ref{Fig: W part, LR}, respectively. It can be seen that the $\boldsymbol{r} \rightarrow \boldsymbol{0}$ limit of the partially intraband potentials is two orders of magnitude larger than the highest energy associated with the long-range (second-order) corrections. Combined with the analogous observations on the fully intraband potentials and the small spatial extent where the second-order corrections are observable, this leads to the conclusion that the long-range second-order corrections are likely negligible for most practical purposes.

\begin{figure}
\centering
\includegraphics[scale=0.75]{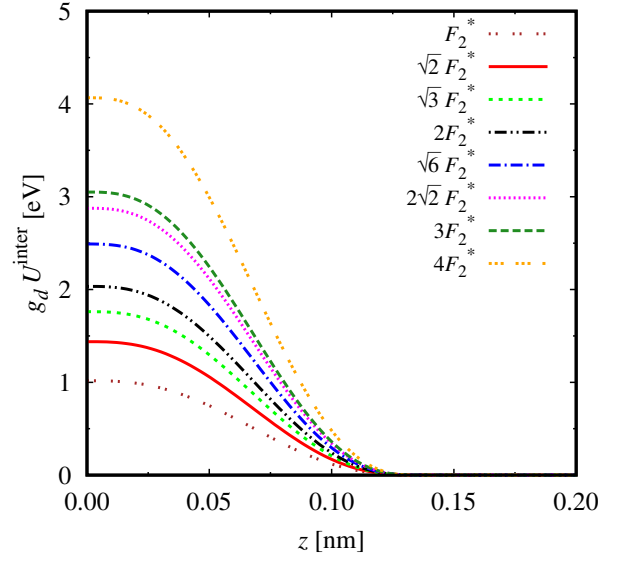}
\caption{ Eight selected interband potentials, plotted along the direction $\boldsymbol{r}  = (0,0,z)$, corresponding to the positive values of $U^{\rm inter}_{\lbrace B \rbrace}$.}
\label{Fig: function gd}
\end{figure}

The interband potentials read as
\begin{align}
\widetilde{W}^{\rm inter}_{\lbrace B \rbrace}(\boldsymbol{r}   )   =  g_d(\boldsymbol{r}  ) \,   U^{\rm inter}_{ \lbrace B \rbrace }      \,,
\label{interband W continuum}
\end{align}
and they are completely short-ranged. In Fig.~\ref{Fig: function gd} we plot 8 such potentials along the $z$ direction, corresponding to the 8 distinct positive values of the parameters $U^{\rm inter}_{\lbrace B \rbrace}$ (see Tables \ref{tab:WSR 1}, \ref{tab:WSR 2} and \ref{tab:WSR 3}). 

We emphasize that the relevance of interband and partially intraband potentials needs to be assessed according to their effect on the envelope functions. Indeed, despite their smaller energy scale with respect to fully intraband processes, interband transitions represent new channels for band mixing, whose effect might possibly be comparable to that of the magnetic field and spin-orbit coupling in strongly confined systems, such as quantum dots. In the context of Si-based quantum computing, where small amounts of band mixing can significantly affect the qubit functionalities, these contributions should also be included. 

\begin{figure}
\centering
\includegraphics[scale=0.75]{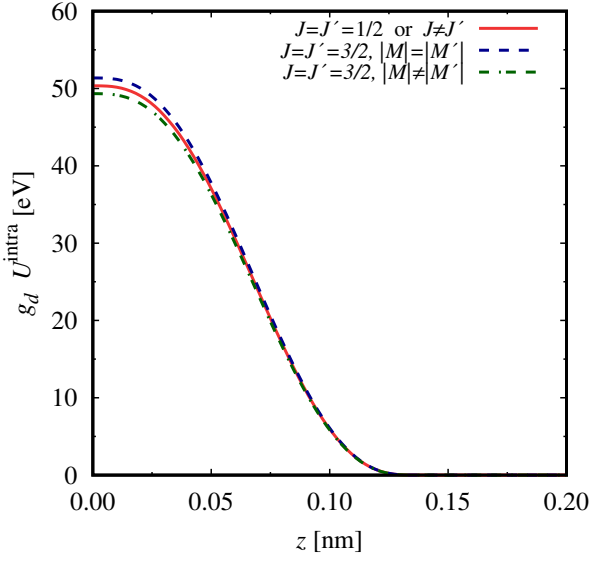}
\caption{Short-range contribution to the intraband potential $\widetilde{W}_{B,B',B',B}(\boldsymbol{r})$, along the direction $\boldsymbol{r}   = (0,0,z)$. The energy splitting of the potentials corresponding to different values of $B=(J,M)$ and $B'=(J',M')$ is apparent close to $\boldsymbol{r} = \boldsymbol{0}$. The three distinct values at $\boldsymbol{r}= \boldsymbol{0}$ are $F_0$ and $F_0 \pm F_2^*$, according to Eq.~\eqref{USR fully intra}.}
\label{Fig: W intra, SR}
\end{figure}

\begin{figure}
\centering
\includegraphics[scale=0.75]{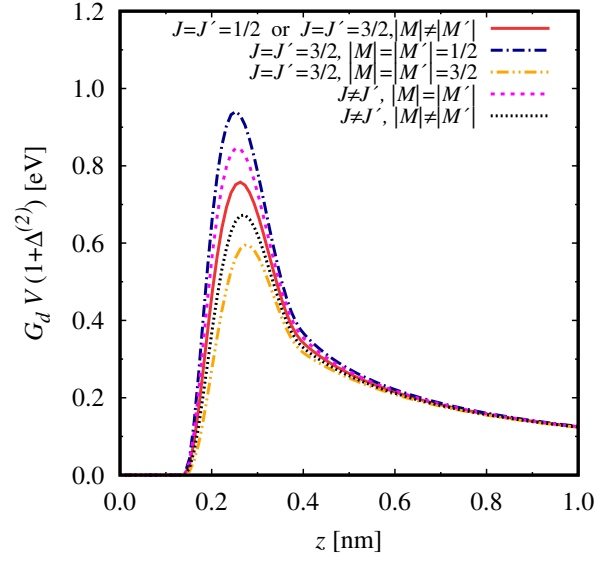}
\caption{ Long-range contribution to the intraband potential $\widetilde{W}_{B,B',B',B}(\boldsymbol{r} )$, along the direction $\boldsymbol{r}  = (0,0,z)$. The differences in the curves are due to the different values taken by $\Delta^{(2)}_{B,B'}$ for different values of $B$ and $B'$. Compare with Fig.~\ref{Fig: W intra, second order}. }
\label{Fig: W intra, LR}
\end{figure}

\begin{figure}
\centering
\includegraphics[scale=0.75]{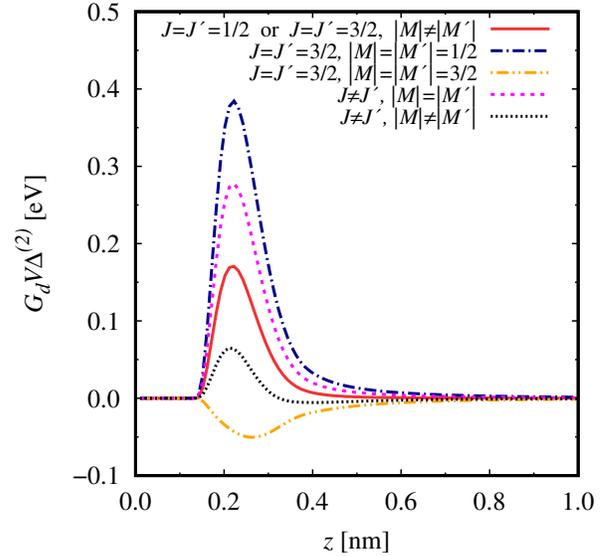}
\caption{ Long-range second-order correction to the full intraband potential $\widetilde{W}_{B,B',B',B}(\boldsymbol{r} )$  in Eq.~\eqref{intraband W continuum} along the direction $\boldsymbol{r}   = (0,0,z)$, labelled by the band indexes $B=(J,M)$ and $B'=(J',M')$. }
\label{Fig: W intra, second order}
\end{figure}

\begin{figure}
\centering
\includegraphics[scale=0.75]{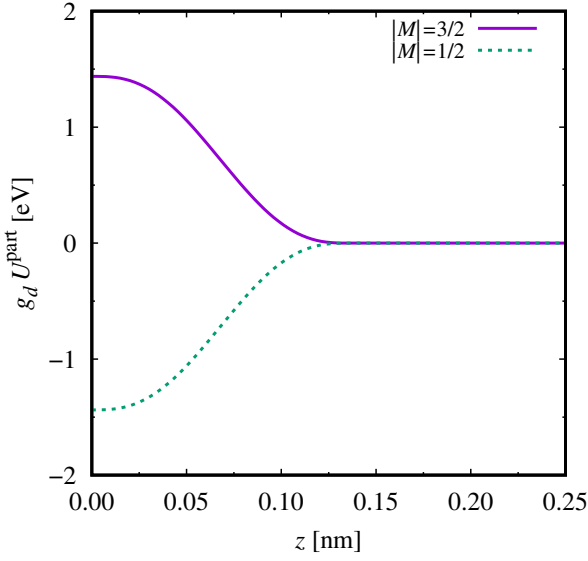}
\caption{ Short-range contribution to the partially intraband potential $\widetilde{W}_{B,B',B",B}(\boldsymbol{r} )$  in Eq.~\eqref{partially intraband W continuum} along the direction $\boldsymbol{r}   = (z,0,z)$, labelled by the band indexes, $B=(J,M)$ and $B'=(J',M')$.}
\label{Fig: W part, SR}
\end{figure}

\begin{figure}
\centering
\includegraphics[scale=0.75]{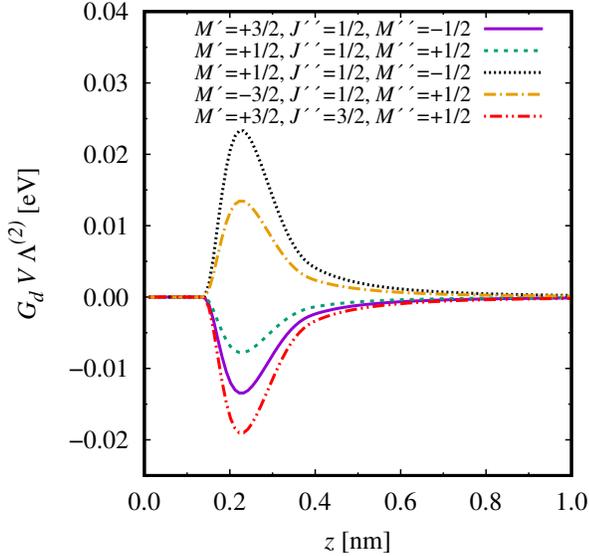}
\caption{ Long-range contribution to the partially intraband potential $\widetilde{W}_{B,B',B",B}(\boldsymbol{r} )$  in Eq.~\eqref{partially intraband W continuum} along the direction $\boldsymbol{r}  = (z,0,z)$ for several selected transitions with $J'=3/2$.}
\label{Fig: W part, LR}
\end{figure}

\section{Numerical results}
\label{sec: Numerical}

In order to illustrate the impact of interactions in Si QDs and quantify the role of short-range interactions, we now present the results of configuration-interaction calculations of the two-hole energy eigenvalues for three prototypical (harmonic) confinements. For a spatially slowly-varying confinement potential $V(\boldsymbol{r})$, the single-hole Hamiltonian is written according to the L\"uttinger-Kohn $\boldsymbol{k} \cdot \boldsymbol{p}$ formula \cite{LK, Voon_book, Secchi21},
\begin{align}
H_{\rm LK} = H_{\boldsymbol{k} \cdot \boldsymbol{p}} + {\rm diag}\left[ V_{\rm QD}(\boldsymbol{r}) \right] \,,
\end{align}
where $H_{\boldsymbol{k} \cdot \boldsymbol{p}}$ is the 6-band $\boldsymbol{k} \cdot \boldsymbol{p}$ kinetic-energy operator, and 
\begin{align}
V_{\rm QD}(\boldsymbol{r}) = \frac{1}{2} \left( \kappa_x x^2 + \kappa_y y^2 + \kappa_z z^2\right)
\end{align}
is a 3D harmonic potential, which models an anisotropic single QD confinement. Rather than to the spring constants $\kappa_{\alpha}$, with $\alpha \in \lbrace x,y,z \rbrace$, in the following we refer to the characteristic confinement lengths $\ell_{\alpha} = \sqrt{ \hbar \gamma_1 / ( m_0 \omega_{\alpha} ) }$, where $\gamma_1 = 4.285$ is the first L\"uttinger parameter for Si, $m_0$ is the bare electron mass, and $\omega_{\alpha} = \sqrt{\kappa_{\alpha} \gamma_1 / m_0} $.

We consider three QDs (QD1, QD2, and QD3), specified by the following confinement lengths $\boldsymbol{\ell} =(\ell_x, \ell_y, \ell_z)$:
\begin{align}
& {\rm QD1:} \quad \boldsymbol{\ell}_1 = (20, 2, 2) \,\, {\rm nm} \,, \nonumber \\
& {\rm QD2:} \quad \boldsymbol{\ell}_2 = (10, 4, 2) \,\, {\rm nm} \,, \nonumber \\
& {\rm QD3:} \quad \boldsymbol{\ell}_3 = (4, 4, 4) \,\, {\rm nm} \,.
\label{definition QDs}
\end{align} 
The characteristic energy scale associated with a harmonic confinement is the effective frequency $\omega^*_{\alpha} = \sqrt{ \kappa_{\alpha}  / m^* }$, where $m^*$ is the effective mass of the confined particles. In a multiband system, the definition of the effective mass is not trivial; we consider here $m^* = \left( \gamma_1 + \frac{5}{2} \gamma_2 \right) m_0$, which is the isotropic part of the effective mass tensor for the light/heavy-hole subsystem \cite{Voon_book}. The energy quanta $\hbar \omega^*_{\alpha}$ corresponding to the considered values of $\ell_{\alpha}$ are:
\begin{align}
& \ell_{\alpha} = 2 \,\, {\rm nm} \,\,\, \Rightarrow  \hbar \omega^*_{\alpha} = 89.337 \,\, {\rm meV} \,, \nonumber \\
& \ell_{\alpha} = 4 \,\, {\rm nm} \,\,\, \Rightarrow  \hbar \omega^*_{\alpha} = 22.334 \,\, {\rm meV} \,, \nonumber \\
& \ell_{\alpha} = 10 \,\, {\rm nm} \,\,\, \Rightarrow \hbar \omega^*_{\alpha} = 3.573 \,\, {\rm meV} \,, \nonumber \\
& \ell_{\alpha} = 20 \,\, {\rm nm} \,\,\, \Rightarrow \hbar \omega^*_{\alpha} = 0.893 \,\, {\rm meV} \,.
\label{table omega}
\end{align}
For each of the cases listed in \eqref{definition QDs}, we study the impact of interactions (both short- and long- ranged) on the two-hole eigenvalues. The latter are obtained from the exact numerical diagonalization of the two-hole Hamiltonian, according to the general procedure that we have presented in Ref.~\onlinecite{Secchi21} for the study of double QDs. In that case, however, interband Coulomb interactions could be neglected because the two holes tend to localize in different dots, so their distance is always very large with respect to the typical range of interband interactions, which are all short-ranged. Since here we consider single QDs, with different confinement strengths, we include all the interaction processes derived above.

As can be expected, long- and short- range Coulomb interactions are in competition, and their interplay is affected by the strength of the confinement potential. Two qualitative pictures can be considered as a reference: 1) when the confinement is relatively weak, the long-range Coulomb repulsion causes the particles to localize far away from each other, forming a Wigner molecule (WM): in these situations the short-range interactions are completely negligible; 2) when the confinement is relatively strong, the two holes are constrained to be close to each other near the center of the QD, despite the Coulomb repulsion: in these situations, short-range interactions can play a role, which we quantify in the following.

Although the interaction potential derived by Vinsome and Richardson is appropriate for isolated bulk Si, we remark that, in a real device, the Si QD is embedded in a dielectric environment which can screen the long-range repulsion between holes. The precise form of the dielectric function is then device-specific. For the sake of generality, we here use the following form of the dielectric function,
\begin{align}
\epsilon(r) = \left\{ \begin{matrix}   \left( \epsilon_0 - 1 \right) \frac{ r }{r_0} + 1 \quad {\rm for} \quad r \leq r_{0} \\
 \epsilon_0 \quad {\rm for} \quad r > r_{0} \end{matrix} \right.  \,,
\end{align} 
which is very similar to the Vinsome-Richardson formula when the parameter $r_0 = 0.3$ nm, and it allows to perform a systematic study of the dependence of the eigenvalues on the screening. In particular, we study the evolution of the two-hole eigenvalues as a function of the parameter $1/\epsilon_0$ ranging from 0 (which suppresses the long-range interaction) to 0.0855, corresponding to the inverse dielectric constant of isolated bulk Si. Since the screening induced by the dielectric environment always increases the long-range screening, $1/\epsilon_0$ can never be higher than 0.0855 in a pure Si QD, and the range $[0, 0.0855]$ covers all possible cases.

\begin{figure}
\centering
\includegraphics[scale=1]{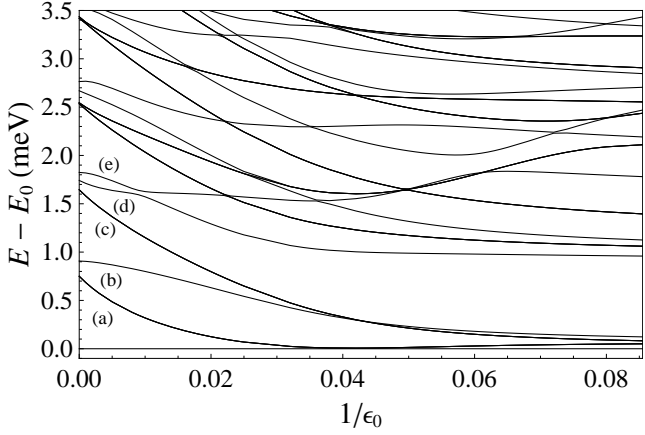}
\caption{Two-hole excitation energies for QD1 [see Eqs.~\eqref{definition QDs}], as functions of $1/\epsilon_0$. The ground state is a singlet; among the curves labelled with letters in the plot, (a) and (c) are triplets, while (b), (d) and (e) are singlets.}
\label{Fig: 2h_20_02_02}
\end{figure}

\begin{figure}
\centering
\includegraphics[scale=1]{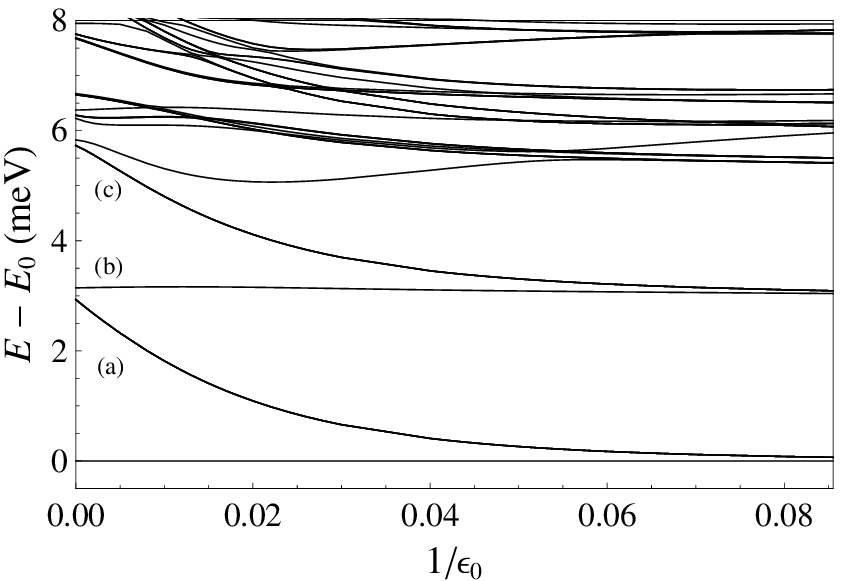}
\caption{Two-hole excitation energies for QD2 [see Eqs.~\eqref{definition QDs}], as functions of $1/\epsilon_0$. The ground state is a singlet; among the curves labelled with letters in the plot, (a) and (c) are triplets, while (b) is a singlet.}
\label{Fig: 2h_10_04_02}
\end{figure}

\begin{figure}
\centering
\includegraphics[scale=1]{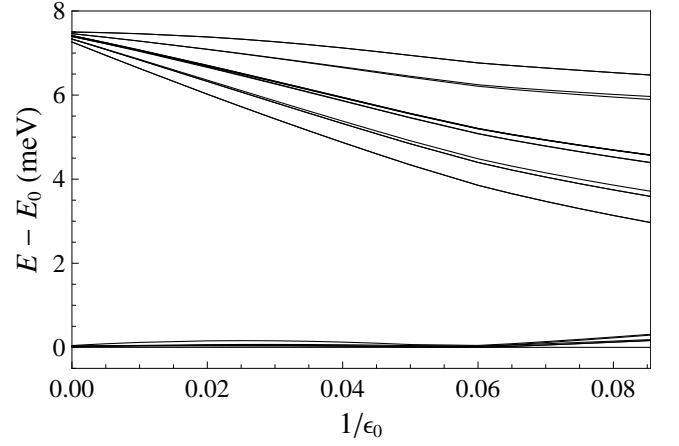}
\caption{Two-hole excitation energies for QD3 [see Eqs.~\eqref{definition QDs}], as functions of $1/\epsilon_0$. The ground manifold is made of 6 states, split by interactions on the scale of $10^{-1}$ meV in the fully interacting regime (see Table \ref{tab: impact SR QD3}).}
\label{Fig: 2h_04_04_04}
\end{figure}

\begin{table}
\centering
 \begin{tabular}{|c||c|c||c|c|} 
 \hline
   & only SR & non interacting &  fully interacting  \\ [0.5ex] 
 \hline 
 $\Delta E_{\rm a}$  & $0.750$ & $0.900$ & $0.051$  \\  [0.5ex]
 $\Delta E_{\rm b}$  & $0.907$ & $0.900$ & $0.123$  \\ [0.5ex]
 $\Delta E_{\rm c}$  & $1.645$ & $1.794$ & $0.082$  \\ [0.5ex]
 $\Delta E_{\rm d}$  & $1.741$ & $1.794$ & $0.959$  \\ [0.5ex]
 $\Delta E_{\rm e}$  & $1.824$ & $1.799$ & $1.126$  \\ [0.5ex] 
 \hline
\end{tabular}
\caption{Selected two-hole excitation energies (in meV) of QD1 [compare with Fig.~\ref{Fig: 2h_20_02_02}], computed using the Vinsome-Richardson formula for the dielectric function. Here, $\Delta E_{x} = E_x - E_0$, where $x \in \lbrace {\rm a,b,c,d,e } \rbrace$ and $E_0$ is the ground energy.}
\label{tab: impact SR QD1}
\end{table}

\begin{table}
\centering
 \begin{tabular}{|c||c|c||c|c|} 
 \hline
   & only SR & non interacting &  fully interacting   \\ [0.5ex] 
 \hline 
 $\Delta E_{\rm a}$  & $2.931$ & $3.139$ & $0.066$  \\  [0.5ex]
 $\Delta E_{\rm b}$  & $3.144$ & $3.139$ & $3.041$  \\ [0.5ex]
 $\Delta E_{\rm c}$  & $5.729$ & $5.841$ & $3.091$  \\ [0.5ex]
 \hline
\end{tabular}
\caption{Selected two-hole excitation energies (in meV) of QD2 [compare with Fig.~\ref{Fig: 2h_10_04_02}], computed using the Vinsome-Richardson formula for the dielectric function. Here, $\Delta E_{x} = E_x - E_0$, where $x \in \lbrace {\rm a,b,c }  \rbrace$ and $E_0$ is the ground energy.}
\label{tab: impact SR QD2}
\end{table}

In general, interactions reduce the the two-hole excitation energies with respect to the non-interacting regime. The physical interpretation of the excitations changes from the progressive occupation of single-particle excited states (in the non-interacting case) to the vibrations of the charges around their classical equilibrium positions (in the Wigner-molecule case). The transition is a continuous one, since the system has a finite size.

\subsection{Quantum dot 1}

QD1 is a quasi-1D system, because the confinement is much stronger in the $(y,z)$ plane than in the $x$ direction. The smallest energy scale associated with confinement is $\hbar \omega^*_x = 0.893$ meV. The two-hole excitation energies are displayed in Fig.~\ref{Fig: 2h_20_02_02}. Tracking the evolution of the eigenvalues as functions of $1/\epsilon_0$, we see that
\begin{itemize}
\item At $1/\epsilon_0 = 0$, the excitation energies $\Delta E_x$, for the excitations $x \in \lbrace {\rm a,b,c,d,e} \rbrace$, can be approximately grouped into a quartet [made of triplet (a) and singlet (b)] and a quintet [made of triplet (c) and singlets (d) and (e)]. The internal splittings within the quartet and the quintet are due to the SR interactions (both intra- and inter-band), which are not suppressed by setting $1/\epsilon_0 = 0$. In the non-interacting system, the internal splittings within the quartet and the quintet vanish (compare the first two columns of Table \ref{tab: impact SR QD1}). This shows that, in a regime of high screening, SR interactions can be relevant; e.g., $\Delta E_{\rm b} - \Delta E_{\rm a} = 0.157$ meV at $1/\epsilon_0 = 0$.       
\item Neglecting the splitting due to SR interactions (see the second column of Table \ref{tab: impact SR QD1}), the energy separations between consecutive low-energy multiplets in the non-interacting regime are $\Delta E_{\rm a} = 0.900$ meV and $\Delta E_{\rm c} - \Delta E_{\rm b} = 0.894$ meV, which are compatible with $\hbar \omega^*_x$.
\item  As the long-range interaction is switched on ($1/\epsilon_0$ increases), the singlet, the quartet, and three among the quintet states converge towards a common energy (apart from residual exchange interactions), while two of the quintet states join other, higher-energy states to form excited interacting multiplets. This reorganization of the spectrum when the interaction is fully switched on, and in particular the formation of highly degenerate manifolds, is a typical signature of the formation of a quasi-1D WM \cite{Secchi10, Corrigan21, Ercan21}. The gap between the two lowest manifolds in the fully interacting regime ($1/\epsilon_0 = 0.0855$) is given by $\Delta E_{\rm d} - \frac{\Delta E_{\rm a} + \Delta E_{\rm b} + \Delta E_{\rm c} }{4} = 0.895$ meV (see the third column of Table \ref{tab: impact SR QD1}), which is compatible with $\hbar \omega^*_x$, pointing to a center-of-mass excitation of the quasi-1D WM. 
\item At $1/\epsilon_0 = 0.0855$, intra-band interactions are negligible, i.e., the modification of the energy gaps in the third column of Table \ref{tab: impact SR QD1} when the intra-band terms are set to zero is $< 1 \mu$eV. This is consistent with the WM picture, because the intra-band interactions are all short-ranged and, therefore, are not expected to contribute significantly when the holes are localized far apart. 
\end{itemize}

We notice that the evidence for WM formation in two-{\it electron} QDs in Si heterostructures has been reported in recent experimental and theoretical works \cite{Corrigan21, Ercan21}.

\subsection{Quantum dot 2}
  
We now consider the two-hole spectrum of QD2. In this case, the confinement is still stronger in the $(y,z)$ plane than in the $x$ direction, but the symmetry has been lowered ($\ell_y \neq \ell_z$) and $\ell_x$ has been decreased with respect to QD1. The excitation energies are shown in Fig.~\ref{Fig: 2h_10_04_02}. As in the case of QD1, we notice a characteristic reorganization of the spectrum as the LR interactions are turned on, which points to the formation of a WM. The degeneracies of the WM manifolds are smaller than in the case of QD1 because of the lower symmetry of the confinement potential. The excitation energy of the fifth eigenstate, $\Delta E_{\rm b}$, evolves from $3.144$ meV (at $1/\epsilon_0 = 0$) to $3.041$ meV (at $1/\epsilon_0 = 0.0855$), remaining always close to (but significantly smaller than) $\hbar \omega^*_x = 3.573$ meV [see \eqref{table omega}]. In the fully interacting regime, this is reminiscent of a single center-of-mass excitation of a WM (inspection of the eigenstate confirms that it evolves from essentially a single Slater determinant of single-hole states at $1/\epsilon_0 = 0$, to a strongly correlated state at $1/\epsilon_0 = 0.0855$). However, the dependence of $\Delta E_{\rm b}$ on $1/\epsilon_0$, which would not occur in a one-band harmonic dot, signals a significant interplay between different vibrational modes induced by the non-triviality of the 6-band kinetic-energy operator. The excitation energies $\Delta E_{\rm a}$ and $\Delta E_{\rm c}$ of the two triplets singled out in Fig.~\ref{Fig: 2h_10_04_02} drop from $2.931$ meV to $0.066$ meV, and from $5.729$ meV to $3.091$ meV $\approx \hbar \omega^*_x$, respectively, when moving from the fully screened to the fully interacting regime (see Table \ref{tab: impact SR QD2}). 

Analogously to the case of QD1, also in QD2 we observe that SR interactions are significant in the regime of high screening (e.g., $\Delta E_{\rm b} - \Delta E_{\rm a} = 0.213$ meV at $1/\epsilon_0 = 0$), while they are negligible at $1/\epsilon_0 = 0.0855$, consistently with the interpretation of the unscreened spectrum in terms of the formation of a WM.

\subsection{Quantum dot 3}

Finally, the two-hole spectrum of QD3 does not show any sign of the formation of a WM. In fact, in this case, confinement is very strong along all directions, overcoming the localizing effect of the Coulomb repulsion. The two-hole excitation energies of QD3 are shown in Fig.~\ref{Fig: 2h_04_04_04}. The 6-fold quasi-degeneracy of the ground state manifold, independent of the strength of the LR interaction, is due to the high symmetry of the confinement potential. The SR interactions lift this degeneracy, which would be exact in the completely non-interacting regime. In the fully screened regime, $1/\epsilon_0 = 0$, the degeneracy is lifted by the SR interactions on the scale of $\approx 10^1 \mu$eV (see first column of Table \ref{tab: impact SR QD3}). In this case, the LR interactions are responsible for a larger lifting, as shown in the second column of Table \ref{tab: impact SR QD3} for the fully interacting case. In this regime, the impact of interband interactions is on the scale of a few $\mu$eV up to $\approx 15 \mu$eV for the 5th excited state, as can be seen from the comparison between the second and the third columns of Table \ref{tab: impact SR QD3}; the third column shows the excitation energies obtained when only the intraband interactions (both LR and SR) are included in the calculations.

\begin{table}
\centering
 \begin{tabular}{|c||c||c|c|c|} 
 \hline
    & only SR &    fully interacting & only intraband  \\ [0.5ex] 
 \hline 
 $\Delta E_{1}$  & $0.014$ & $0.167$ & $0.165$  \\  [0.5ex]
 $\Delta E_{2}$  & $0.014$ & $0.167$ & $0.165$  \\ [0.5ex]
 $\Delta E_{3}$  & $0.014$ & $0.189$ & $0.189$  \\ [0.5ex]
 $\Delta E_{4}$  & $0.040$ & $0.299$ & $0.298$  \\  [0.5ex]
 $\Delta E_{5}$  & $0.042$ & $0.319$ & $0.304$  \\ [0.5ex]
 \hline
\end{tabular}
\caption{Excitation energies (in meV) of the first 5 two-hole excited states above the ground state in QD3 [compare with Fig.~\ref{Fig: 2h_04_04_04}], computed using the Vinsome-Richardson formula for the dielectric function. When all interactions are neglected, all these gaps $\Delta E_x = 0$.}
\label{tab: impact SR QD3}
\end{table}

\section{Conclusions}
\label{sec: conc}

In conclusion, we have thoroughly investigated the band scattering processes induced by the Coulomb interaction in a system of holes at the $\boldsymbol{\Gamma}$ point in Si, and derived the relevant potentials. In particular, a set of many previously overlooked interband and partially intraband processes has been derived, most of which are relevant at short length scales. Corrections to the long-range effective interaction, which is usually assumed to be a simple Coulomb intraband potential, have also been derived. Such corrections decay to zero quickly with the inter-hole distance. 

We have performed CI calculations of the two-hole spectra in three exemplary QDs, including all interaction terms, in order to study the impact of long-range and short-range interactions. These calculations show that two holes embedded in realistic QDs in Si tend to form Wigner molecules, whose signature can be seen in the values and degeneracies of the excitation energies. A similar result had been reported for electrons in Si \cite{Corrigan21, Ercan21} but, to the best of our knowledge, not yet for holes. In our numerical calculations, the long-range interaction is gradually switched on by changing the value of the bulk dielectric function from the fully screened regime ($1/\epsilon_0 = 0$) to the fully interacting regime ($1/\epsilon_0 = 0.0855$, the value for isolated bulk Si). Therefore, these calculations should qualitatively reproduce the spectra which can be obtained in the presence of a variable dielectric environment surrounding the Si QDs (e.g., that provided by close metallic gates). The impact of short-range interactions on two-hole spectra is found to be relevant mostly in the regime of high screening due to the dielectric environment, while it becomes essentially negligible in an isolated Si QD.

\acknowledgments
The authors acknowledge financial support from the European Commission through the project IQubits (Call: H2020-FETOPEN-2018-2019-2020-01, Project ID: 829005). The authors acknowledge CINECA for HPC computing resources and support under the ISCRA initiative (IsC87 ESQUDO - HP10CXQWD5).

\appendix

\section{Transformations involving the Clebsch-Gordan coefficients}
\label{app: Clebsch}

The Clebsch-Gordan coefficients appearing in Eq.~\eqref{Bloch states silicon} are:
\begin{align}
& S_{B,\alpha,\sigma} \nonumber \\
& = \delta_{J, \frac{3}{2} } \Bigg[ \delta_{M, \sigma \frac{3}{2}}   \frac{1}{\sqrt{2}} (\delta_{\alpha, x} + {\rm i} \sigma \delta_{\alpha, y})   
 - \delta_{M, \sigma \frac{1}{2}} \sqrt{\frac{2}{3}}   \delta_{\alpha, z} \nonumber \\
& \quad   - \delta_{M, - \sigma \frac{1}{2}}   \frac{   \sigma}{\sqrt{6}}  \left(\delta_{\alpha, x}  -  {\rm i} \sigma \delta_{\alpha, y} \right)  \Bigg] \nonumber \\
& \quad + \delta_{J, \frac{1}{2}} \frac{1}{\sqrt{3}}  \Big[   \delta_{M, \sigma \frac{1}{2}} \delta_{\alpha , z }  
  - \sigma \delta_{M, - \sigma \frac{1}{2}}   \left( \delta_{\alpha , x } - {\rm i}  \sigma \delta_{\alpha , y } \right)   \Big] \,  .  
\label{S matrix 6 bands} 
\end{align}

In the derivation of the effective band-dependent interaction potential, after substituting the expressions of the single-hole eigenstates $|\nu\rangle$ [Eq. \eqref{single-particle eigenstates}] into Eq.~\eqref{Coulomb integrals}, we obtain Eq.~\eqref{band dependent V}, with
\begin{align}
W_{ \lbrace B \rbrace }&\left( \boldsymbol{r} - \boldsymbol{r}' \right) \equiv  V\left( \boldsymbol{r} - \boldsymbol{r}' \right) \frac{1}{ \mathcal{N}^2 }  \sum_{ \lbrace \boldsymbol{R} \rbrace } \sum_{ \lbrace j \rbrace } (-1)^{j_1+j_2+j_3+j_4} \nonumber \\
& \quad \times \sum_{ \lbrace \alpha \rbrace }    
\sum_{\sigma,\sigma' } \left(  S^*_{B_1 , \alpha_1, \sigma } S_{B_4 , \alpha_4, \sigma } \right) 
\left( S^*_{B_2 , \alpha_2, \sigma'} S_{B_3 , \alpha_3, \sigma'} \right)  \nonumber \\
& \quad \times \phi^*_{p_{\alpha_1}}(\boldsymbol{r} - \boldsymbol{R}_{1, j_1}) \, \phi^*_{p_{\alpha_2}}(\boldsymbol{r}' - \boldsymbol{R}_{2, j_2})  \nonumber \\
& \quad \times \phi_{p_{\alpha_3}}(\boldsymbol{r}' - \boldsymbol{R}_{3, j_3}) \,  \phi_{p_{\alpha_4}}(\boldsymbol{r} - \boldsymbol{R}_{ 4 , j_4}) \,.
\label{band dependent interaction}
\end{align}
Eq.~\eqref{band dependent interaction} includes summations having the general form
\begin{align}
\sum_{  \sigma} \left( \sum_{\alpha'} \phi^*_{p_{\alpha'}}(\boldsymbol{x}')     S^*_{B', (\alpha', \sigma )} \right) \left( \sum_{\alpha} S_{B, (\alpha, \sigma )}   \phi_{p_{\alpha}}(\boldsymbol{x}) \right) \,,
\label{interesting sum without localization}
\end{align}
where $\boldsymbol{x}$ and $\boldsymbol{x}'$ denote, in general, two different positions. We perform the summation in Eq.~\eqref{interesting sum without localization}, using Eq.~\eqref{S matrix 6 bands}, and expressing the result in terms of the orbitals given in Eq.~\eqref{orbitals m}. We obtain that the term involving the sum over $\alpha$ in Eq.~\eqref{interesting sum without localization} is
\begin{align}
\sum_{\alpha} S_{B, (\alpha, \sigma )}   \phi_{p_{\alpha}}(\boldsymbol{r})  
& =    
\delta_{J, \frac{3}{2} } \delta_{M,  \frac{3 \sigma}{2}}   \phi_{\sigma}(\boldsymbol{r})  + Y_J \, \delta_{M,  \frac{\sigma}{2}} \phi_{0}(\boldsymbol{r})    \nonumber \\
& \quad     -  \sigma  X_J \, \delta_{M, -  \frac{\sigma}{2}}  \phi_{-\sigma}(\boldsymbol{r}) \,,
\label{sum alpha done}
\end{align}
where $\sigma \in \lbrace +1,-1\rbrace$, and
\begin{align}
& X_J \equiv \frac{1}{\sqrt{3}} \left( \sqrt{2} \delta_{J, \frac{1}{2}} + \delta_{J, \frac{3}{2}} \right) \,, \nonumber \\ 
& Y_J \equiv \frac{1}{\sqrt{3}}  \left( \delta_{J, \frac{1}{2}}   - \delta_{J, \frac{3}{2} }   \sqrt{2}  \right) \,.
\end{align}
The term involving the sum over $\alpha'$ from Eq.~\eqref{interesting sum without localization} is obtained by taking the complex conjugate of Eq.~\eqref{sum alpha done} and changing the indices. We finally write
\begin{align}
& \sum_{\alpha', \alpha, \sigma} \phi^*_{p_{\alpha'}}(\boldsymbol{x}')    S^*_{B', (\alpha', \sigma )} S_{B, (\alpha, \sigma )}   \phi_{p_{\alpha}}(\boldsymbol{x}) \nonumber \\
& \equiv \sum_{m',m} \phi^*_{m'}(\boldsymbol{x}') \, F^{m', m}_{B',B} \, \phi_{m}(\boldsymbol{x}) \,,
\label{introduction F matrix}
\end{align}
where we have introduced the matrix $F^{m', m}_{B',B}$, with elements
\begin{align}
F_{B',B}^{\pm 1, \pm 1}   = \Big( \delta_{J', \frac{3}{2} } \,    \delta_{J, \frac{3}{2} } \, \delta_{M, \pm  \frac{3 }{2}}    + X_{J'} \,  X_J  \,  \delta_{M, \pm   \frac{1}{2}} \Big) \delta_{M', M} \,,
\end{align}
\begin{align}
F_{B',B}^{\pm 1,0}   =   \delta_{J', \frac{3}{2} } \, \delta_{M', \pm \frac{3 }{2}} \,  Y_J \,     \delta_{M, \pm \frac{1}{2}}      \pm  X_{J'} \, \delta_{M', \pm \frac{ 1}{2}} \, Y_J  \,      \delta_{M,  \mp \frac{ 1}{2}} \,,
\end{align}
\begin{align}
F_{B',B}^{0, \pm 1}   = Y_{J'} \,   \delta_{M', \pm \frac{1}{2}}  \,  \delta_{J, \frac{3}{2} } \, \delta_{M, \pm \frac{3}{2}}      \pm Y_{J'}  \, \delta_{M',  \mp \frac{1}{2}} \, X_J \, \delta_{M, \pm \frac{1}{2}} \,, 
\end{align}
\begin{align}
F_{B',B}^{\pm 1, \mp 1}  & = \mp \,  \delta_{J', \frac{3}{2} } \, \delta_{M', \pm \frac{3}{2}} \, X_J \, \delta_{M, \mp  \frac{1}{2}}       \nonumber \\
& \quad  \pm   X_{J'} \, \delta_{M', \pm \frac{1}{2}} \,   \delta_{J, \frac{3}{2} } \, \delta_{M,  \mp \frac{3 }{2}}  \,,
\end{align}
\begin{align}
F_{B',B}^{0,0}   =  Y_{J'} \, Y_J \,  \delta_{M',M}  \, \delta_{| M'| ,  \frac{1}{2}}       \,.
\label{F00}
\end{align}
The matrix $F$ has the following trace property:
\begin{align}
  \sum_{m} F^{  m    ,   m}_{B',B} =  \delta_{B', B }           \,,
\label{trace F matrix}
\end{align}
as can be easily derived after noticing that
\begin{align}
X_{J'} \, X_J + Y_{J'} \, Y_J = \delta_{J', J} \,.
\end{align}
By applying Eq.~\eqref{introduction F matrix} to Eq.~\eqref{band dependent interaction}, we obtain Eq.~\eqref{W before localization}.

\section{Screened potential}
\label{app: screened potential}

We here discuss the explicit expression of the screened Coulomb potential $V(\boldsymbol{r} - \boldsymbol{r}')$ in Si, which enters Eq.~\eqref{W before localization}. 
The screened Coulomb interaction potential between two holes at distance $r$ reads   
\begin{align}
V( r  ) = \frac{ e^2}{\epsilon(r) \, r} \equiv \frac{V_{\rm C}(r)}{\epsilon(r)} \,, 
\label{screened Coulomb}
\end{align}
where $\epsilon(r)$ is a static, isotropic\cite{Phillips61} but nonhomogeneous dielectric function for Si, and $V_{\rm C}(r) = e^2 / r $ is the bare Coulomb potential.

The modelling of $\epsilon(r)$ has a long history\cite{BassaniIadonisiPreziosi}, from the semiclassical Thomas-Fermi theory\cite{Resta77, Chandramohan90, Franceschetti05}, to quantum-mechanical models with simplifying assumptions on the band structure\cite{Penn62, Srinivasan68}, to more refined numerical calculations based on empirical pseudo-potential methods\cite{Nara65, Nara66, Vinsome71, Richardson71}. The numerical approaches account for material-specific details, such as the crystal band structure and the correct electronic dispersion, thus allowing for a description of the optical properties of materials which is more precise than that provided by analytical models. Moreover, they predict the bulk value $\epsilon_0 \equiv \epsilon(r \rightarrow \infty)$ of the dielectric function. 

\begin{figure}
\centering
\includegraphics[scale=0.75]{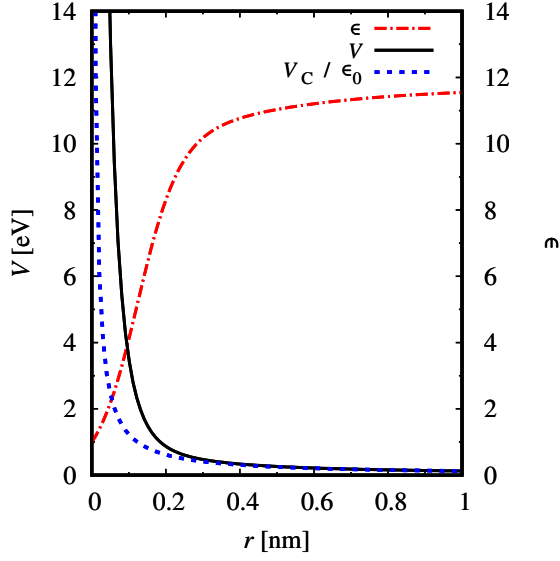}
\caption{Left vertical axis: Screened potential $V(r)$ with $\epsilon(r)$ given by Eq.~\eqref{epsilon(r) VR} (solid black curve), and  Coulomb potential screened at all $r$ by the bulk dielectric constant $\epsilon_0$ (dashed blue curve), plotted for comparison. Right vertical axis: Dielectric function $\epsilon(r)$, according to Eq.~\eqref{epsilon(r) VR}.}
\label{Fig: Potential VR}
\end{figure}

Numerical calculations are usually supplemented by interpolation functions, and thus lead to analytical expressions for $\epsilon(r)$. We take as a reference the works of Vinsome and Richardson\cite{Vinsome71, Richardson71} on a comprehensive set of zincblende semiconductors. They perform large-scale RPA calculations (in reciprocal space), and they interpolate their numerical results with the following formula, valid in direct space:
\begin{align}
 \epsilon(r)  =  \left( \frac{1}{\epsilon_0} + \lambda_1 {\rm e}^{- 2 \pi \alpha_1 r / a } + \lambda_2 {\rm e}^{- 2 \pi \alpha_2 r/a}   \right)^{-1}   \,, 
 \label{epsilon(r) VR}
\end{align}
where $a$ is the cubic cell edge, and the fitting parameters for Si are written as\cite{Richardson71}
\begin{align}
& \frac{1}{\epsilon_0} \equiv \frac{B}{D} \,, \quad \quad \alpha_{1,2} \equiv \left( \frac{C \mp \sqrt{C^2 - 4 D}}{2} \right)^{1/2} \,,  \nonumber \\ 
& \lambda_{1,2} \equiv \frac{1}{2} \left( 1 - \frac{1}{\epsilon_0} \right) \pm \frac{A - \frac{C}{2} \left( 1 + \frac{1}{\epsilon_0} \right)}{\sqrt{C^2 - 4 D}} \,,
\end{align}
in terms of the quantities
\begin{align}
A = 0.34 \,, \quad B = 0.016 \,, \quad C = 2.6 \,, \quad D = 0.17 \,.
\label{VR parameters}
\end{align}
The bulk limit for the dielectric function according to Eq.~\eqref{VR parameters} is $\epsilon_0 = 10.625$. By modifying the value of $B$ to $0.01453$, one obtains $\epsilon_0 = 11.7$, consistently with experimental data\cite{Dunlap53}. The potential $V(r)$ is plotted in Fig.~\ref{Fig: Potential VR}.

The explicit expressions for the quantities $\mathcal{L}^{-2}(r)$ and $\mathcal{M}^{-2}(r)$, contributing to the second-order correction to the long-range effective potentials [see Eqs.~\eqref{Lm2 general} and \eqref{Mm2 general}], are 
\begin{align}
  \mathcal{M}^{-2}(r)   =    \frac{1}{ r^2}     +  \frac{\epsilon(r)}{r }    \frac{2 \pi  }{a  }      \sum_{n=1}^2 \lambda_n \alpha_n {\rm e}^{-   2 \pi \alpha_n   r / a }    \,,
  \label{Mcal}
\end{align}
and
\begin{align}
 \mathcal{L}^{-2}(r)   \equiv  3 \mathcal{M}^{-2}(r)   + \epsilon(r) \frac{4 \pi^2 }{a^2 }  \sum_{n=1}^2 \lambda_n \alpha^2_n  {\rm e}^{-   2 \pi \alpha_n   r / a } \,.  
\label{Lcal}
\end{align}

\section{R\"udenberg approach for multi-center integrals}
\label{app: Rudenberg}

A possible way to go beyond the two-center integral approximation can be outlined as follows. Following R\"udenberg\cite{Rudenberg51}, we introduce a complete set of orthonormal orbitals centered at each atomic site $\boldsymbol{R}_j$:
\begin{align}
\Big\{ \big| \chi , \boldsymbol{R}_j \big> \,, \quad   \chi = 1s, \ldots 3p_{1}, 3p_0, 3p_{-1},   \ldots \Big\} \,.
\end{align}
We define the overlap between orbitals centered at different atomic sites:
\begin{align}
\mathcal{O}_{\chi  \, , \, \chi'  }\left( \boldsymbol{R}_j, \boldsymbol{R}'_{j'} \right) \equiv  \big<  \chi, \boldsymbol{R}_j  \big|     \chi',  \boldsymbol{R}'_{j'} \big> \,.
\end{align}
Due to the completeness of the set centered at any arbitrary site, we can expand the orbital centered at one site in terms of the orbitals centered at another site. According to R\"udenberg, at least if $\boldsymbol{R}$ and $\boldsymbol{R}'$ are very close, it can be assumed that the only relevant contribution in the expansion of an orbital $\phi_{m}(\boldsymbol{r} - \boldsymbol{R}_{j'}')$ is the one from   $\phi_{m}(\boldsymbol{r} - \boldsymbol{R}_{j})$.

A possible way to refine this approximation is to consider instead the full set of $3p$ orbitals (i.e., we allow $\chi$ to be equal not just to $m'$, but to any of the basis orbitals). Extending this to arbitrary atomic positions, one obtains for the interaction potential [Eq.~\eqref{W before localization}] the following expression:
\begin{align}
& W_{\lbrace M \rbrace }\left( \boldsymbol{r} - \boldsymbol{r}' \right) \nonumber \\
& \approx  V\left( \boldsymbol{r} - \boldsymbol{r}' \right) \frac{1}{ \mathcal{N}^2 }  \sum_{   \boldsymbol{R}  , \boldsymbol{R}'  } \sum_{  j , j' }    \sum_{ \lbrace m \rbrace}    
         \Delta^{m_1, m_4}_{M_1, M_4}\left( \boldsymbol{R}_{  j }   \right) \nonumber \\ 
 & \quad \times  \Delta^{m_2, m_3}_{M_2, M_3}\left( \boldsymbol{R}'_{j'}   \right) \,  \phi^*_{m_1}(\boldsymbol{r} - \boldsymbol{R}_{ j }) \, \phi^*_{m_2}(\boldsymbol{r}' - \boldsymbol{R}'_{  j' }) \nonumber \\
& \quad \times \phi_{m_3}(\boldsymbol{r}' - \boldsymbol{R}'_{  j' }) \, \phi_{m_4}(\boldsymbol{r} - \boldsymbol{R}_{ j })  \,.
\label{band dependent interaction - first approximation}
\end{align}
where we have introduced the overlap form factor,
\begin{align}
& \Delta^{m_1, m_4}_{M_1, M_4}\left( \boldsymbol{R}_{  j }    \right) \nonumber \\
& \equiv \frac{1}{2} \sum_{\boldsymbol{R}'', j''} (-1)^{j + j''} \sum_{m_5} \Big[ F_{M_1 , M_4}^{m_5, m_4} \mathcal{O}_{m_5 , m_1 }\left( \boldsymbol{R}''_{j''}, \boldsymbol{R}_{  j } \right) \nonumber \\
& \quad +  F_{M_1 , M_4}^{m_1, m_5} \mathcal{O}_{m_4, m_5 }\left( \boldsymbol{R}_{  j }, \boldsymbol{R}''_{j''} \right) \Big] \,.
\label{Delta overlap}
\end{align}
The R\"udenberg approximation is recovered by assuming
\begin{align}
\mathcal{O}_{m_4, m_5 }\left( \boldsymbol{R}_{  j }, \boldsymbol{R}''_{j''} \right)  \approx  \delta_{m_4, m_5} \mathcal{O}_{m_4, m_4 }\left( \boldsymbol{R}_{  j }, \boldsymbol{R}''_{j''} \right) \,,
\end{align}
which yields
\begin{align}
& \Delta^{m_1, m_4}_{M_1, M_4}\left( \boldsymbol{R}_{  j }    \right) \nonumber \\
&  \approx  F_{M_1 , M_4}^{m_1, m_4} \frac{1}{2} \sum_{\boldsymbol{R}'', j''} (-1)^{j + j''}   \Big[  \mathcal{O}_{m_1,  m_1 }\left( \boldsymbol{R}''_{j''}, \boldsymbol{R}_{  j } \right) \nonumber \\
& \quad +    \mathcal{O}_{m_4, m_4 }\left( \boldsymbol{R}_{  j }, \boldsymbol{R}''_{j''} \right) \Big] \,.
\label{Delta overlap Rudenberg}
\end{align}
The two-center integral approximation is formally recovered by replacing
\begin{align}
\mathcal{O}_{m, m} \left( \boldsymbol{R}_j , \boldsymbol{R}'_{j'} \right) \rightarrow \delta_{\boldsymbol{R} , \boldsymbol{R}' }  \delta_{j,j'} \,,
\end{align} 
which yields
\begin{align}
\Delta^{m_1, m_4}_{M_1, M_4}\left( \boldsymbol{R}_{  j }    \right) \rightarrow  F_{M_1 , M_4}^{m_1, m_4} \,.
\end{align}

Thus, we have seen that a possible strategy for improving over the two-center approximation requires the calculation of the quantity \eqref{Delta overlap}. We notice that, in a lattice, 
\begin{align}
\mathcal{O}_{\chi, m' }\left(\boldsymbol{R}_{j} , \boldsymbol{R}'_{j'}  \right)  =  \mathcal{O}_{\chi, m' }\left(\boldsymbol{R}  - \boldsymbol{R}' ;  j -  j'   \right) \,,
\end{align}
therefore, $\Delta^{m_1, m_4}_{M_1, M_4}\left( \boldsymbol{R}_{  j }    \right)$ is actually independent of $\boldsymbol{R}_{j} $:
\begin{align}
& \Delta^{m_1, m_4}_{M_1, M_4} \nonumber \\
& \equiv \frac{1}{2} \sum_{  j''} (-1)^{j + j''} \sum_{m_5} \Big\{ F_{M_1 , M_4}^{m_5, m_4} \sum_{\boldsymbol{R}'' } \mathcal{O}_{m_5 , m_1 }\left( \boldsymbol{R}'' ; j''-j \right) \nonumber \\
& \quad +  F_{M_1 , M_4}^{m_1, m_5} \sum_{\boldsymbol{R}'' } \mathcal{O}_{m_4, m_5 }\left(   \boldsymbol{R}''  ; j - j'' \right) \Big\} \,.
\label{Delta overlap in a lattice}
\end{align}

\section{Derivation of the Hubbard parameters}
\label{app: Hubbard}

\subsection{General remarks}

We rewrite Eq.~\eqref{Hubbard parameters} here in a more general way as
\begin{align}
    U_{i,j,k,l} & = \int d\boldsymbol{r}_1 \int d\boldsymbol{r}_2 \, \phi^*_{i}(\boldsymbol{r}_1)\phi^*_{j}(\boldsymbol{r}_2)V_{\rm C}(|\boldsymbol{r}_1 - \boldsymbol{r}_2 |) \nonumber \\
    & \quad \times \phi_{k}(\boldsymbol{r}_2)\phi_l(\boldsymbol{r}_1) \,,
     \label{eq:onectwoeint}
\end{align}
where the atomic orbital $\phi_i$ is separable into the product of a radial part and a spherical harmonic,
\begin{align}
    \phi_i(\boldsymbol{r})= R_{n_i , l_i}(r) \, \Theta_{l_i, m_i}(\theta) \, \Phi_{m_i}(\varphi)  \, ,
\end{align}
where \cite{Slater29}
\begin{align}
    \Phi_{m}(\varphi) = \frac{1}{\sqrt{2\pi}} {\rm e}^{{\rm i } m\varphi} \,, \label{eq:phim}
\end{align}
\begin{align}
    \Theta_{l, m}(\theta) = \sqrt{\frac{(2l+1)}{2}\frac{(l-|m|)!}{(l+|m|)!}}P_{l, |m|}(\cos\theta) \,, \label{eq:thetalm}
\end{align}
\begin{align}
    P_{l, |m|}(\cos\theta) = \frac{1}{2^ll!}\sin^{|m|}\theta\frac{d^{|m|+l}(-\sin^2\theta)^l}{d(\cos\theta)^{|m|+l}} \,.
    \label{Plm}
\end{align}
In the definition of the spherical harmonics, we have followed the convention adopted in Ref.~\onlinecite{Voon_book}, i.e. the Condon-Shortley phase $(-1)^{m}$ for $m\geq 0$ is {\it not} included.

We then use the expansion of the Coulomb potential in series of Legendre polynomials, Eq.~\eqref{Legendre expansion}. After substituting it into Eq.~\eqref{eq:onectwoeint}, and performing some standard manipulations that involve the spherical harmonic addition theorem\cite{Griffith}, we obtain
\begin{align}
    U_{i,j,k,l} & = \delta_{m_i + m_j , m_k + m_l} \sum_{\ell=0}^{\infty} R_{\ell}(n_i, l_i ; n_j, l_j ; n_k, l_k; n_l, l_l) \nonumber \\
    & \quad \times c_{\ell}(l_i, m_i; l_l, m_l) \, c_{\ell}(l_k, m_k ; l_j, m_j) \,,
    \label{U in terms of c}
\end{align}
where
\begin{align}
&     R_{\ell}\left(n_i, l_i; n_j, l_j ; n_k, l_k ; n_l, l_l \right) \nonumber \\
& = e^2\int_0^{\infty} dr_1 r_1^2 \int_{0}^{\infty} dr_2 r_2^2 \, R_{n_i, l_i}(r_1)R_{n_j, l_j}(r_2)\frac{r_<^{\ell}}{r_>^{\ell+1}}   \nonumber \\
& \quad \times R_{n_k, l_k}(r_2)R_{n_l, l_l}(r_1) \,,
\end{align}
and
\begin{align}
    c_{\ell}(l,m;l',m')  
    &= \sqrt{\frac{2}{2\ell+1}} \int_{0}^{\pi}  d\theta \sin(\theta) \, \Theta_{l,m}(\theta) \nonumber \\
    & \quad \times \Theta_{\ell, m-m'}(\theta) \, \Theta_{l',m'}(\theta) \,.
    \label{definition c}
\end{align}
This quantity vanishes unless 
\begin{align}
\ell+l+l' \,\, {\rm is \,\, even} \quad \land \quad |l-l'|\leq\ell \leq l+l' \,.
\label{important condition}
\end{align}

\subsection{Valence orbitals in Silicon}
 
In this work, we need only considering the case of  
\begin{align}
n_i=n_j=n_k=n_l=3 \,, \quad  l_i=l_j=l_k=l_l=1 \,,
\end{align}
since we are only concerned with $3p$ atomic orbitals. From the condition Eq.~\eqref{important condition}, we then see that the only nonvanishing terms in Eq.~\eqref{U in terms of c} are those with $\ell \in \lbrace 0,2 \rbrace$.

Since $n$ and $l$ are fixed, we restore the notation of Eq.~\eqref{Hubbard parameters}, where only the $m$ numbers are specified explicitly. Analogously, we put $c_{\ell}(1, m ; 1, m') \equiv c_{\ell}(m , m')$. We also introduce the Slater-Condon parameters
\begin{align}
F_0(3p,3p) & \equiv F_0  \equiv R_0(3,1;3,1;3,1;3,1) \nonumber \\
&  = e^2 \! \int_0^{\infty} dr_1 r_1^2 \int_0^{\infty} dr_2 r_2^2 \,  \frac{1}{r_>} \, R^2_{3,1}(r_1)R^2_{3,1}(r_2) \,, \nonumber \\
F_2(3p,3p) & \equiv F_2 \equiv R_2(3,1;3,1;3,1;3,1) \nonumber \\
& = e^2 \! \int_0^{\infty} dr_1 r_1^2 \int_0^{\infty} dr_2 r_2^2 \,  \frac{r^2_<}{r^3_>}R^2_{3,1}(r_1)R^2_{3,1}(r_2) \,,
\label{Slater-Condon generic}
\end{align} 
where $r_<=\text{min}(r_1,r_2)$ and $r_>=\text{max}(r_1,r_2)$. The quantities \eqref{Slater-Condon generic} coincide with those introduced in Eq.~\eqref{Slater-Condon}. The Hubbard parameters in Eq.~\eqref{Hubbard parameters} are then reduced to
\begin{align}
    U_{\lbrace m \rbrace} & = \delta_{m_1+m_2,m_3+m_4} \Big[ F_0  \, c_0(m_1 , m_4) \, c_0(m_3,m_2) \nonumber \\
    & \quad + F_2  \, c_2(m_1,m_4) \, c_2(m_3,m_2) \Big] \,.
    \label{Hubbard in terms of c and F}
\end{align}

We evaluate $c_0(m,m')$ and $c_2(m,m')$ analytically, using Eqs.~\eqref{eq:thetalm} and \eqref{Plm}. We obtain
\begin{align}
c_0(m,m') = \delta_{m,m'}   \,,  
\label{c0 result}
\end{align}
and
\begin{align}
c_2(m,m') & = \delta_{m,m'}  \frac{(-1)^{|m|} \left( 2 - |m| \right) }{5} \nonumber \\
& \quad + (1 - \delta_{m,m'}) \frac{\sqrt{3 \left(|m| + |m'|\right)} }{5}   \,.
\label{c2 result} 
\end{align}
After substituting Eq.~\eqref{c2 result} into Eq.~\eqref{Hubbard in terms of c and F}, we obtain Eq.~\eqref{Hubbard in terms of F}.

\section{Derivation of the short-range potentials}
\label{app: short-range}

Using Eqs.~\eqref{independent Hubbard}, we rewrite Eq.~\eqref{W SR} as
\begin{align}
& W^{\rm SR}_{\lbrace B \rbrace }(\boldsymbol{R}_j, \boldsymbol{R}'_{j'} )  
\nonumber \\
& =  \delta_{\boldsymbol{R}_j, \boldsymbol{R}'_{j'}} \Big\{     F^{ 0, 0}_{ B_1 , B_4 } \, F^{ 0, 0}_{ B_2 ,  B_3 } U_{0, 0, 0, 0} \nonumber \\
& \quad 
 +  \Big( F^{ 1, 1}_{ B_1 ,  B_4 }   + F^{ -1, -1}_{ B_1 ,  B_4 } \Big) \Big( F^{ 1, 1}_{ B_2 ,  B_3 }       +  F^{ -1, -1}_{ B_2 ,  B_3 }     \Big) U_{1, 1, 1, 1}  
 \nonumber \\
& \quad + \Big[  \Big( F^{ 1, 1}_{ B_1 , B_4 }  
+F^{ -1, -1}_{ B_1 ,  B_4 } \Big) F^{ 0, 0}_{ B_2 ,  B_3 } \nonumber \\
& \quad
+F^{ 0, 0}_{ B_1 , B_4 } \Big( F^{ 1, 1}_{ B_2 ,  B_3 }
+  F^{ -1, -1}_{ B_2 ,  B_3 } \Big) \Big] U_{1, 0, 0, 1} \nonumber \\
& \quad + \Big( F^{ 1, -1}_{ B_1 ,  B_4 } \, F^{ -1, 1}_{ B_2 ,  B_3 } 
 + F^{-1, 1}_{ B_1 ,  B_4 } \, F^{ 1, -1}_{ B_2 ,  B_3 } \Big) U_{1, -1, 1, -1} \nonumber \\
& \quad + \Big[ \Big( F^{ 0, -1}_{ B_1 ,  B_4 } + F^{ 1, 0}_{ B_1 ,  B_4 } \Big)  \Big( F^{ 0, 1}_{ B_2 ,  B_3 }  + F^{ -1, 0}_{ B_2 ,  B_3 }   \Big) \nonumber \\
& \quad 
+ \Big( F^{ 0, 1}_{ B_1 ,  B_4 } +   F^{ -1, 0}_{ B_1 , B_4 }  \Big) \Big( F^{ 0, -1}_{ B_2 , B_3 } + F^{ 1, 0}_{ B_2 ,  B_3 } \Big)  \Big] U_{0, 0, 1, -1} \Big\}  \,.
\label{W SR before analytical manipulation}
\end{align}
We use Eq.~\eqref{independent Hubbard}, we rearrange some terms, and use Eq.~\eqref{trace F matrix}, and we obtain
\begin{align}
& W^{\rm SR}_{\lbrace B \rbrace }(\boldsymbol{R}_j, \boldsymbol{R}_{j} )  
\nonumber \\
& =      F^{  0, 0}_{  B_1 ,   B_4 } \, F^{  0, 0}_{  B_2 ,   B_3 } \, 9  F^{\star}_2    
 +  \delta_{B_1, B_4} \delta_{  B_2 ,   B_3 }   \left(  F_0    +     F^{\star}_2 \right)  
 \nonumber \\
& \quad -  \delta_{  B_1 ,   B_4 }  
  F^{ 0, 0}_{ B_2 ,   B_3 }  3 F^{\star}_2     - F^{  0, 0}_{  B_1 ,   B_4 } \delta_{B2, B3} \, 3  F^{\star}_2      \nonumber \\
& \quad + \Big( F^{\, 1, -1}_{\, B_1 , \, B_4 } \, F^{\, -1, 1}_{\, B_2 , \, B_3 } 
 + F^{\, -1, 1}_{\, B_1 , \, B_4 } \, F^{\, 1, -1}_{\, B_2 , \, B_3 } \Big) 6  F^{\star}_2 \nonumber \\
& \quad + \Big[ \Big( F^{ 0, -1}_{ B_1 ,  B_4 } + F^{ 1, 0}_{ B_1 ,  B_4 } \Big)  \Big( F^{ 0, 1}_{ B_2 ,  B_3 }  + F^{ -1, 0}_{ B_2 ,  B_3 }   \Big) \nonumber \\
& \quad 
+ \Big( F^{ 0, 1}_{ B_1 ,  B_4 } +   F^{ -1, 0}_{ B_1 , B_4 }  \Big) \Big( F^{ 0, -1}_{ B_2 , B_3 } + F^{ 1, 0}_{ B_2 ,  B_3 } \Big)  \Big] 3  F^{\star}_2   \,.
\label{W SR manipulation in progress}  
\end{align}
Let us examine the various scattering processes implied by Eq.~\eqref{W SR manipulation in progress}. The first two lines involve the matrix element $F^{0,0}_{B',B}$, which we can rewrite from Eq.~\eqref{F00} as
\begin{align}
F_{B',B}^{0,0} 
& =   \frac{1}{3}  \left[ \delta_{J', J} \left( J' + \frac{1}{2} \right)   - \sqrt{2} \left(1 - \delta_{J',J} \right) \right]   \nonumber \\
& \quad \times  \delta_{M',M}  \, \delta_{| M'| ,  \frac{1}{2}}   \,.
\end{align}
It can be seen that $F_{B',B}^{0,0}$ provides a term which conserves the band ($\propto \delta_{J',J} \delta_{M',M}$) and a term which induces a transition between bands [$\propto \left(1 - \delta_{J',J} \right) \delta_{M',M}$] at one of the interaction vertices. The various combinations appearing in the first two lines of the right-hand side of Eq.~\eqref{W SR manipulation in progress}, therefore, include intraband, partially intraband, and interband processes. On the other hand, the last three lines correspond to interband scattering processes. The latter involve combinations of the form
\begin{align}
& F^{ 0, -1}_{ B' ,  B  } + F^{ 1, 0}_{ B' ,  B  } =  F^{ -1, 0}_{ B ,  B'  } + F^{ 0, 1}_{ B ,  B'  } \nonumber \\
& = Y_{J'}   \,  \delta_{J, \frac{3}{2} } \,   \delta_{M', - \frac{1}{2}}  \, \delta_{M, - \frac{3}{2}}       +  \delta_{J', \frac{3}{2} }  \,  Y_J  \, \delta_{M',  \frac{3 }{2}} \,     \delta_{M,   \frac{1}{2}}  \nonumber \\
& \quad       + \left( J'-J \right)  \delta_{M',   \frac{ 1}{2}}   \,      \delta_{M,  - \frac{ 1}{2}} \,,
\end{align}
where we have used the relation
\begin{align}
X_{J'} \, Y_J  - Y_{J'}    \, X_J = J' - J \,,
\end{align}
valid for $J, J' \in \lbrace 3/2, 1/2 \rbrace$.

Making all terms explicit, we separate Eq.~\eqref{W SR manipulation in progress} as in Eq.~\eqref{WSR separated}, with the three individual terms given by Eqs.~\eqref{WSR fully intra}, \eqref{WSR partial}, and \eqref{WSR inter}.

The fully intraband potential is given by
\begin{align}
& U_{B_1,B_2}^{\rm intra} \nonumber \\
& =  F_0 +  F^{\star}_2 \Bigg[  1  -   \left( J_1 + \frac{1}{2} \right) \delta_{|M_1|, \frac{1}{2}} -   \left( J_2 + \frac{1}{2} \right) \delta_{|M_2|, \frac{1}{2}} \nonumber \\
& \quad   + \left( J_1 + \frac{1}{2} \right) \delta_{|M_1|, \frac{1}{2}}  \left( J_2 + \frac{1}{2} \right) \delta_{|M_2|, \frac{1}{2}}     \Bigg] \,.
\end{align}
After a few algebraic manipulations and making use of the fact that $J = 1/2 \Rightarrow |M| = 1/2$, this expression can be shown to be equivalent to Eq.~\eqref{USR fully intra} of the main text.

The partially intraband potential is given by
\begin{align}
U^{\rm part}_{B_1; B_2, B_3} & = \left[ 1 -      \left( J_1 + \frac{1}{2} \right) \delta_{|M_1|, \frac{1}{2}} \right]  \sqrt{2} \left( 1 - \delta_{J_2, J_3}\right)     
 \nonumber \\
& \quad \times \delta_{M_2, M_3} \delta_{|M_2|, \frac{1}{2}}    F^{\star}_2  \,.  
\end{align}
In a similar way to the previous case, this expression can be shown to be equivalent to Eq.~\eqref{USR partial} of the main text.

The completely interband potential is separated into two parts: the first one originates as a part of the term $F^{  0, 0}_{  B_1 ,   B_4 } \, F^{  0, 0}_{  B_2 ,   B_3 } \, 9  F^{\star}_2  $ of Eq.~\eqref{W SR manipulation in progress}, and is directly given by Eq.~\eqref{USR inter 1} of the main text. The second one originates from the last three lines of Eq.~\eqref{W SR manipulation in progress}, and is given by 
\begin{align}
& U^{(2),\,\rm inter}_{B_1, B_4; B_2, B_3} \nonumber \\
& \equiv  \Big( F^{ 0, -1}_{ B_1 ,  B_4 } + F^{ 1, 0}_{ B_1 ,  B_4 } \Big)  \Big( F^{ 0, 1}_{ B_2 ,  B_3 }  + F^{ -1, 0}_{ B_2 ,  B_3 }   \Big)  3  F^{\star}_2 \nonumber \\
& \quad + F^{  1, -1}_{  B_1 ,   B_4 } \, F^{  -1, 1}_{  B_2 ,  B_3 } \, 6 F_2^{\star} \,,
\end{align}
which is turned into Eq.~\eqref{USR inter 2} after some algebraic passages. Note that a term $U^{(2),\,\rm inter}_{B_2, B_3; B_1, B_4} $ is also included in Eq.~\eqref{WSR inter}.

\section{$\alpha\beta$-integrals}
\label{app: alphabeta}

We here provide the expressions for the $\alpha\beta$-integrals, introduced in Eq.~\eqref{alphabeta integrals before}, as functions of the quantity \eqref{eq:intrad2}. Using the spherical harmonics introduced in Appendix \ref{app: Hubbard}, from a straightforward integration over the polar angles we obtain 
\begin{align}
 \int {\rm d} \boldsymbol{r}   \, \phi^*_{m }(\boldsymbol{r}  ) \,     x^2     \,   \phi_{m}(\boldsymbol{r} )  =    2^{|m|}   \frac{ 1  }{ 5   }  \left< r^2_{3,1} \right>         \,,
\end{align}
\begin{align}
 \int {\rm d} \boldsymbol{r}   \, \phi^*_{\pm 1}(\boldsymbol{r}  ) \,     x^2     \,   \phi_{\mp 1}(\boldsymbol{r} ) =     \frac{ 1  }{ 5   }  \left< r^2_{3,1} \right>         \,,
\end{align}
\begin{align}
 \int {\rm d} \boldsymbol{r}   \, \phi^*_{m }(\boldsymbol{r}  ) \,     y^2     \,   \phi_{m}(\boldsymbol{r} )  =    2^{|m|}   \frac{ 1  }{ 5   }  \left< r^2_{3,1} \right>         \,,
\end{align}
\begin{align}
 \int {\rm d} \boldsymbol{r}   \, \phi^*_{\pm 1}(\boldsymbol{r}  ) \,     y^2     \,   \phi_{\mp 1}(\boldsymbol{r} ) =  - \frac{ 1  }{ 5   }  \left< r^2_{3,1} \right>         \,,
\end{align}
\begin{align}
   \int {\rm d} \boldsymbol{r}   \, \phi^*_{m }(\boldsymbol{r}  ) \,     z^2     \,   \phi_{m'}(\boldsymbol{r} )    = \delta_{m, m'} \frac{  3^{1 - |m|}}{5 }       \left< r^2_{3,1} \right>          \,,
\end{align}
\begin{align}
   \int {\rm d} \boldsymbol{r}   \, \phi^*_{\pm 1}(\boldsymbol{r}  ) \,     x y     \,   \phi_{\mp 1}(\boldsymbol{r} )    =     \mp     \frac{   {\rm i}   }{  5   }   \left< r^2_{3,1} \right>    \,,
\end{align}
\begin{align}
  \int {\rm d} \boldsymbol{r}   \, \phi^*_{0}(\boldsymbol{r}  ) \,     y z     \,   \phi_{\pm 1}(\boldsymbol{r} )   = \pm \frac{   {\rm i} }{ 5 \sqrt{ 2}  }     \left< r^2_{3,1} \right>      \,, 
\end{align}
\begin{align}
 \int {\rm d} \boldsymbol{r}   \, \phi^*_{0}(\boldsymbol{r}  ) \,     z x     \,   \phi_{\pm 1}(\boldsymbol{r} )  =   \frac{ 1 }{ 5  \sqrt{ 2 } }     \left< r^2_{3,1} \right>        \,.
\end{align}

\section{Evaluation of $\left< r^2_{3,1}\right>$}
\label{app: r2}

In order to compute $W^{\rm{ LR, (2)}}_{\lbrace B \rbrace}(\boldsymbol{R})$, we need to evaluate the integral in Eq.~\eqref{eq:intrad2} analytically and numerically. This task requires the choice of a specific form for the radial wave functions associated with the tight-binding orbitals. We present and compare two different approaches.

\subsection{Hartree-Fock atomic orbitals}

We first compute $\left< r^2_{3,1} \right>$ using the Hartree-Fock (HF) radial orbitals as provided by Watson and Freeman\cite{Watson61}. They compute the atomic orbitals for neutral Silicon $\left( 1s^2 2s^2 2p^6 3s^2 3p^2  ,  \, ^3P \right)$ by applying the variational principle to the total energy of the system, where the many-electron Hamiltonian for Si atom contains the kinetic energy, nuclear potential energy and inter-electronic electrostatic energy. Within their method they assume that there is only one radial wave function per shell, which is the average of those corresponding to the different occupied orbitals of that shell. 

In particular, the radial wave function for the $3p$ shell is written as a linear combination of Slater-type radial orbitals $R_i(\rho)$, with $\rho = r / a_{\rm B}$:
\begin{align}
    U_{3p}(\rho)=\sum_{i}C^{3p}_{i}R_i(\rho) \,,
    \label{Watson Freeman U3p}
\end{align}
where 
\begin{align}
    R_i(\rho)=\sqrt{\frac{(2Z_i)^{5+2A_i}}{(4+2A_i)!}}\rho^{2+A_i} {\rm e}^{-Z_i\rho} \,;
\end{align}
the normalization is
\begin{align}
    \int_0^{\infty} \left| U_{3p}(\rho) \right|^2 d\rho = 1 \,.
\end{align} 
According to Ref.~\onlinecite{Watson61}, 7 basis function are needed in Eq.~\eqref{Watson Freeman U3p}. For the sake of completeness, in Table \ref{HF parameters} we report the values of the coefficients $A_i$, $Z_i$ and $C_i^{3p}$, taken from Ref.~\onlinecite{Watson61}.

The evaluation of the parameters $F_0$ and $F_2$ using these HF radial functions yields the numerical values given in Eq.~\eqref{numerical values}. Using the same radial functions, we evaluate
\begin{align}
    \left<r^2_{3,1}\right>^{\rm (HF)}   =  \int_0^{\infty}|U_{3p}(\rho)|^2\rho^2d\rho   =0.0252 \, {\rm nm}^2 \,. 
    \label{r2 HF estimate}
\end{align}

\begin{table}
\centering
 \begin{tabular}{|c c c c|} 
 \hline
 $i$ & $A_i$ & $Z_i$ & $C_i^{3p}$ \\ [0.5ex] 
 \hline 
 1 & 0 & $10.8139$ & $-0.01181046$ \\  [0.5ex]
 2 & 0 & $6.8493$ & $-0.03787150$ \\ [0.5ex]
 3 & 0 & $4.2336$ & $-0.17923597$ \\ [0.5ex]
 4 & 1 & $3.3949$ & $0.02649990$ \\ [0.5ex]
 5 & 1 & $1.7195$ & $0.34702725$ \\ [0.5ex] 
 6 & 1 & $1.1824$ & $0.63306352$ \\ [0.5ex] 
 7 & 1 & $0.5932$ & $0.08747425$ \\ [0.5ex] 
 \hline
\end{tabular}
\caption{Parameters of the HF radial wave functions \cite{Watson61}.}
\label{HF parameters}
\end{table}

\subsection{Hydrogen-ion atomic orbitals}
We now derive $\left< r^2_{3,1} \right>$, as well as $F_0$ and $F_2$, using hydrogen-ion (HI) atomic orbitals, whose radial wave function is
\begin{align}
    R_{n,l}( r)&=\sqrt{\left(\frac{2Z^{\star}}{na_{\rm B}}\right)^3\frac{(n-l-1)!}{2n(n+l)!}} \, \exp\left(-\frac{ Z^{\star} r }{n a_{{\rm B}}} \right)  \nonumber \\
    & \quad \times \left(\frac{2Z^{\star} r}{na_{\rm B}}\right)^l L^{2l+1}_{n-l-1}\!\left(\frac{2Z^{\star}r}{na_{\rm B}}\right)\,,
\end{align}
where $L^{2l+1}_{n-l-1}(x)$ is a generalized Laguerre polynomial, $Z^{\star}$ is an effective screened nuclear charge, and $a_{\rm B} =0.05291$ nm. For $n=3$ and $l=1$ the radial orbital reads as
\begin{align}
     R_{3,1}(r)  
       = \left. \frac{1}{9\sqrt{6}}\left(\frac{Z^{\star}}{a_{\rm B}}\right)^{3/2} {\rm e}^{-x/2}x(4-x) \right|_{x =   2Z^{\star} r  / (3a_{\rm B} ) }  \,.
       \label{radial HI}
\end{align}

The attractive feature of Eq.~\eqref{radial HI} is that it depends on a single parameter $Z^{\star}$. We can then evaluate Eq.~\eqref{eq:intrad2}, as well as the Slater-Condon parameters $F_0$ and $F_2$ defined in Eq.~\eqref{Slater-Condon}, and the three resulting formulas will depend only on $Z^{\star}$. We obtain
\begin{align}
    \left< r^2_{3,1} \right>^{\rm (HI)}  
     = 180 \left(\frac{a_{\rm B}}{Z^{\star}} \right)^2  = \frac{ 0.5039 \, {\rm nm}^2 }{\left( Z^{\star} \right)^2 }   \,,
    \label{r2 HI}
\end{align}
\begin{align}
    F^{\rm (HI)}_0  = 0.07186 \frac{Z^{\star}e^2}{a_{\rm B}} = 1.9557 \,{\rm eV} \, \times Z^{\star} \,, 
    \label{F0 HI}
\end{align}
and
\begin{align}
 F^{\rm (HI)}_2   =0.03598 \frac{Z^{\star}e^2}{a_{\rm B}} = 0.9792 \,{\rm eV} \, \times Z^{\star}\,,
 \label{F2 HI}
\end{align}
where we have used $e^2 = 1.4399764  \, {\rm eV} \cdot {\rm nm}$. 

The values of $F_0$ and $F_2$ given in Ref.~\onlinecite{Watson61}, that we reported in Eq.~\eqref{numerical values}, are reproduced by our Eqs.~\eqref{F0 HI} and \eqref{F2 HI} for $Z^{\star} = 4.597$ and $Z^{\star} =4.636$, respectively. Hence, the picture in terms of HI orbitals is compatible with the results of Ref.~\onlinecite{Watson61}, provided that we assume an effective nuclear charge of $Z^{\star} \approx 4.6$. This seems to be consistent with the intuitive picture that little less than 10 core electrons ($n=1$, $n=2$) in a Si atom screen the nucleus charge seen by the electrons in the $3p$ orbitals with respect to the bare nucleus charge $Z = 14$. Using $Z^{\star} = 4.6$, we obtain from \eqref{r2 HI} the estimate
\begin{align}
 \left< r^2_{3,1} \right>^{\rm (HI)}  
      =  0.0238 \, {\rm nm}^2     \,,
    \label{r2 HI estimate}
\end{align}
which is remarkably close to the value obtained using the HF radial orbitals, Eq.~\eqref{r2 HF estimate}.

\section{Smooth functions for the continuum limit of the effective potentials}
\label{app: smooth}
 
\subsection{An exact solution for the $g$ function}

Consider the surface $\mathcal{S}_{\boldsymbol{0}}(L)$ of the cube centered on the origin and with edge $L>0$. Analogously, $\mathcal{S}_{\boldsymbol{R}_j}(L)$ is the surface of edge $L$ centered on the atomic position $\boldsymbol{R}_j$. We look for a function $g(\boldsymbol{r})$ such that
\begin{align}
g(\boldsymbol{r}) \equiv \left\{ \begin{matrix} \eta_0(L) & {\rm if} \,\, \boldsymbol{r} \in \mathcal{S}_{\boldsymbol{0}}(L) \,, & 0   < L  \leq   \lambda  \\ 
0 & {\rm if} \,\, \boldsymbol{r} \in \mathcal{S}_{\boldsymbol{0}}(L) \,, &  L >  \lambda   \end{matrix} \right.  \,\,.
\label{g(r)}
\end{align}
The condition $\boldsymbol{r} \in \mathcal{S}_{\boldsymbol{0}}(L)$ can be translated into
\begin{align}
L \equiv L_{\boldsymbol{r}} \equiv 2 \max(|x|,|y|,|z|) \,, \quad {\rm where} \,\, \boldsymbol{r} \equiv (x,y,z) \,.
\label{L r}
\end{align}
The cubic surfaces $\mathcal{S}_{\boldsymbol{0}}(L)$ are thus isosurfaces of $g(\boldsymbol{r})$, which vanishes outside the cube $\mathcal{C}_{\boldsymbol{0}}$, whose surface is $\mathcal{S}_{\boldsymbol{0}}(\lambda)$. The requirement of continuity of $g(\boldsymbol{r})$ at the surface of $\mathcal{C}_{\boldsymbol{0}}$ and Eq.~\eqref{equal 1} impose the following conditions on $\eta_0(L)$: 
\begin{align}
& \eta_0(\lambda ) = 0 \,, 
&    
\int_{0}^{\lambda }  d L  \,   L^2 \, \eta_0(L) = 1/3 \,.
\label{mandatory eta0}
\end{align}
The latter condition has been derived from: $d \mathcal{V}(L) \equiv \mathcal{V}(L+dL) - \mathcal{V}(L) \approx 3L^2 dL$, where $\mathcal{V}(L)$ is the volume of a cube of edge $L$, and $dL$ is its infinitesimal increment. In addition to these mandatory requirements, we are free to impose conditions of smoothness, such as
\begin{align}
&  \partial_L \eta_0(L) \big|_{L = 0} = 0 \,, \quad 
   \partial^2_{L,L} \eta_0(L) \big|_{L = 0} = 0 \,, \nonumber \\
&  \partial_L \eta_0(L) \big|_{L = \lambda} = 0 \,, \quad 
   \partial^2_{L,L} \eta_0(L) \big|_{L = \lambda} = 0 \,.
\label{optional eta0}   
\end{align}
The lowest-order polynomial function satisfying both Eqs.~\eqref{mandatory eta0} and \eqref{optional eta0} is
\begin{align}
\eta_0(L) = \frac{1}{\lambda^3} \left[ \frac{28}{5} - 56 \left( \frac{L}{\lambda}\right)^3 + 84 \left( \frac{L}{\lambda}\right)^4 - \frac{168}{5} \left( \frac{L}{\lambda}\right)^5 \right] \,.
\label{eta0 solution}
\end{align}

\subsection{A computationally feasible solution}

We assume that
\begin{align}
\widetilde{F}(\boldsymbol{r}) = \eta(L) \quad {\rm if} \,\, \boldsymbol{r} \in \mathcal{S}_{\boldsymbol{0}}(L) \,, \quad \lambda   < L < 3 \lambda  \,,
\end{align}
i.e., that the cubic surfaces $\mathcal{S}_{\boldsymbol{0}}(L)$ are isosurfaces of $\widetilde{F}(\boldsymbol{r})$, outside the cube $\mathcal{C}_{\boldsymbol{0}}$ (where we do not modify $F$). The function $\eta(L)$ must satisfy the following constraints due to continuity:
\begin{align}
\eta(\lambda ) = 0 \,, \quad\quad
\eta(3 \lambda ) = \rho \,,
\end{align}
and the integral constraint:
\begin{align}
  \int_{\mathcal{R} \setminus \mathcal{C}_{\boldsymbol{0}}} d \boldsymbol{r} \, \widetilde{F}(\boldsymbol{r}) = 3 \int_{\lambda }^{3 \lambda }  d L  \,   L^2 \, \eta(L) = 26
\,.
\end{align} 
We also impose the optional smoothness conditions
\begin{align}
& \partial_L \left. \eta(L) \right|_{L = \lambda } = 0 \,, 
 \quad \quad
\partial_L \left. \eta(L) \right|_{L = 3\lambda } = 0 \,, \nonumber \\ 
& \partial^2_{L,L} \left. \eta(L) \right|_{L = \lambda } = 0 \,, \quad \quad \partial^2_{L,L} \left. \eta(L) \right|_{L = 3\lambda } = 0 \,.
\end{align}

The lowest-order polynomial that satisfies all conditions (both mandatory and optional) has the form 
\begin{align}
\eta(L) \equiv \frac{1}{\lambda^3} \sum_{n= 0}^{6} b_n \left( \frac{   L }{\lambda} \right)^n \,,
\label{eta(L)}
\end{align}
with
\begin{align}
& b_0 = -\frac{25087}{1184} \,, 
\quad b_1 =  \frac{49545}{592} \,, 
\quad b_2 =   -\frac{154395}{1184} \,, \nonumber \\
& b_3 =  \frac{30085}{296}  \,, \quad b_4 =  -\frac{49245}{1184} \,, \quad b_5 =  \frac{5061}{592} \,, \nonumber \\ 
& b_6 =  -\frac{825}{1184} \,. 
\label{coefficients of eta(L)}
\end{align}

The effect of using $\widetilde{F}(\boldsymbol{r})$ is that the density of atoms is not equally distributed anymore in the 27 cubes forming $\mathcal{R}$. For the nearest, next-nearest, and next-next-nearest neighbours, we respectively find 
\begin{align}
& \int_{\mathcal{C}_{(1,0,0)}} d \boldsymbol{r} \widetilde{F}(\boldsymbol{r}) =  
\frac{212}{259} \approx 0.8185 \,, \nonumber \\
& \int_{\mathcal{C}_{(1,0,1)}} d \boldsymbol{r} \widetilde{F}(\boldsymbol{r}) =  
\frac{ 535 }{ 518 } \approx 1.0328  \,, \nonumber \\
& \int_{\mathcal{C}_{(1,1,1)}} d \boldsymbol{r} \widetilde{F}(\boldsymbol{r}) =   
\frac{ 563 }{ 518 } \approx 1.0869  \,.
\end{align}

\subsection{Results for the smoothing functions}

To summarize, a solution for the functions $g_d$ and $G_d$ appearing in Eq.~\eqref{W continuum - step 2} is given by
\begin{align}
g_d(\boldsymbol{r}) & \equiv \Theta_{2 L_{\boldsymbol{r}} \leq a} \left[ \frac{28}{5} - 56 \left( \frac{2 L_{\boldsymbol{r}}}{a}\right)^3 + 84 \left( \frac{2 L_{\boldsymbol{r}}}{a}\right)^4 \right. \nonumber \\
& \quad \left. - \, \frac{168}{5} \left( \frac{2 L_{\boldsymbol{r}}}{a}\right)^5 \right]   \,,     
\end{align}
and
\begin{align}
G_d(\boldsymbol{r}) = \left\{ \begin{matrix} 
0 \,, &  0 \leq 2 L_{\boldsymbol{r}} < a  \\
\sum_{n= 0}^{6} b_n \! \left( \frac{ 2  L_{\boldsymbol{r}} }{a} \right)^n \,,  &   a \leq 2 L_{\boldsymbol{r}} \leq  3a  \\  
1 \,, &   2 L_{\boldsymbol{r}} >  3a   \end{matrix} \right. \,,
\end{align}
where the coefficients $b_n$ are given by \eqref{coefficients of eta(L)}.

\end{document}